\documentclass[11pt]{article}
\usepackage{amsmath}
\usepackage{amsthm}
\usepackage{amssymb}
\usepackage{mathtools}
\usepackage[pdftex]{graphicx}
\usepackage[update, prepend, verbose]{epstopdf}
\usepackage{verbatim}
\usepackage{enumerate}
\usepackage{natbib}
\usepackage{textcomp}
\usepackage[capposition=top]{floatrow}
\usepackage{setspace}
\usepackage{ bbold }
\usepackage{bbm}
\usepackage{tikz}
\usepackage[utf8]{inputenc}
\usepackage[T1]{fontenc}
\usepackage[polish,english]{babel}
\usepackage{hyperref}
\usepackage{xr}
\usepackage{booktabs}
\newtheorem{exmp}{Example}
\newtheorem{assumption}{Assumption}
\newtheorem{theorem}{Theorem}
\newtheorem{proposition}{Proposition}
\newtheorem{lemma}{Lemma}

\newtheorem{remark}{Remark}
\newtheorem{customassumption}{Assumption}
\newenvironment{customassmpt}[1]
  {\renewcommand\thecustomassumption{#1}\customassumption}
  {\endcustomassumption}

\def \eps{\varepsilon}

\title{Testing Shape Restrictions with Continuous Treatment: A Transformation Model Approach\footnote{I would like to thank Nikolas Mittag and Pedro Sant'Anna for useful discussions and comments. This research used the ALICE High Performance Computing Facility at the University of Leicester and HPC computing facilities at the University of Kent.}}
\author{Arkadiusz Szydłowski\footnote{School of Economics, Politics and IR, Sibson Building, University of Kent, Canterbury CT2 7PE, UK. \emph{E-mail address}: \href{mailto:a.szydlowski@kent.ac.uk}{a.szydlowski@kent.ac.uk}}\\
\emph{University of Kent} \\
\small
\normalsize
}
\begin{document}
%\begin{spacing}{1.5}
\maketitle

\begin{abstract}
We propose tests for the convexity/linearity/concavity of a transformation of the dependent variable in a semiparametric transformation model. These tests can be used to verify monotonicity of the treatment effect, or, equivalently, concavity/convexity of the outcome with respect to the treatment, in (quasi-)experimental settings. Our procedure does not require estimation of the transformation or the distribution of the error terms. The statistic takes the form of a U statistic or a localised U statistic, and we show that critical values can be obtained by bootstrapping. In our application we test the convexity of loan demand with respect to the interest rate using experimental data from South Africa. \\
\vspace{2cm}
\end{abstract}
%\end{spacing}

\section{Introduction} \label{intro}

In this paper we consider testing if a treatment has a diminishing or increasing effect on the outcome. This is often of interest on top of the question if there is an effect at all or what sign it has. For example, often it is natural to expect that demand will be decreasing in price. However, it is less clear if increasing the price will have a larger effect at lower or higher price levels. Our test can address this question without fully estimating the demand relationship.

Let $X$ denote the vector of treatment and control variables. As estimating the effect of $X$ on the outcome $Y$ nonparametrically  with non-trivial number of treatments and controls would suffer from a curse of dimensionality, we impose a single-index structure and assume that the nonlinearity of the treatment effect comes from a nonlinear transformation of $Y$. In other words, consider a transformation model of the form:
\begin{gather}
T(Y)=X'\beta_0+\varepsilon \label{1}
\end{gather}  
where $Y$ is a scalar dependent variable, $X$ is a vector of $q$ nondegenerate explanatory variables, $\beta_0$ is a vector of coefficients belonging to a compact set $\Theta_{\beta} \subset \mathbb{R}^q$, $T(\cdot)$ is an increasing function and $\varepsilon$ is an unobserved error term with distribution $F$ that is independent of $X$. In order to identify the model we need a location normalisation: e.g. $T(0)=0$, $E(\eps)=0$ or $Me(\eps)=0$; and a scale normalisation: e.g. $|\beta_{0,1}|=1$. The benefit of using the transformation model compared to a standard single-index model (i.e. $Y=T(X'\beta_0) + \varepsilon$), besides the fact that it facilitates our testing approach, is that the transformation model allows the treatment effect of $X_{k}$ to depend on the values of both observed and unobserved characteristics (note that $Y=T^{-1}(X'\beta_0+\varepsilon)$) so can be seen as a simple way of introducing heterogenous treatment effects that vary with unobservables. 

The main objective of this article is to develop a practically appealing test to determine whether the transformation function $T(\cdot)$ is concave/linear/convex. The examples below illustrate the importance of testing curvature of the transformation.
\begin{exmp}\emph{\textbf{(Experiments with continuous treatment)}}
Using normalization $E(\varepsilon) = 0$ and $Q_{\alpha}(\varepsilon) = 0$ where $Q_{\alpha}$ denotes the $\alpha$ quantile, we have, respectively:
\begin{align*}
\frac{\partial^{2} E(Y|X)}{\partial X_{k}^{2}} &= - \beta_{0,k}^{2} E\left[ \frac{T''(T^{-1}(X'\beta_0+\eps))}{T'(T^{-1}(X'\beta_0+\eps))^{3}}\bigg| X\right] \qquad \text{(mean regression)} \\
\intertext{and}
\frac{\partial^{2} Q_{\alpha}(Y|X)}{\partial X_{k}^{2}} &= -\beta_{0,k}^{2} \frac{T''(T^{-1}(X'\beta_0))} {T'(T^{-1}(X'\beta_0))^{3}}
\qquad \text{(quantile regression)} 
\end{align*}
As $T'(\cdot)>0$ the curvature of the mean/quantile effect depends on $T''(\cdot)$. Thus, the sign of the second derivative $T''(\cdot)$ determines if the effect of the treatment $X$ on the $\alpha$ quantile of $Y$ is concave or convex in $X$. For example, if a company randomises marketing spending in different markets, testing for concavity would answer the question if marketing spending has diminishing returns (e.g. on mean revenue).  Test of curvature has also application in experimental studies of demand elasticities (e.g. \cite{jessoe_rapson14}, \cite{hainmueller_et_al15}, \cite{karlan_zinman19}) where it can be used to verify if demand is concave, linear or convex. 

Finally, when the transformation is linear, the treatment effect does not depend on the observed ($X$) and unobserved ($\varepsilon$) heterogeneity. Thus, the test of linearity of $T$ can be seen as a test of treatment effect heterogeneity. 
\end{exmp}

\begin{exmp} \emph{\textbf{(Duration models: testing hazard monotonicity)}}
Let $\lambda(\cdot)$ and $\Lambda(\cdot)$ denote baseline hazard and integrated baseline hazard, respectively. In a duration model: $T(Y) = \log \Lambda (Y)$, and
\begin{gather*}
T''(Y) =  \frac{\lambda'(Y)}{\Lambda(Y)}-\left(\frac{\lambda(Y)}{\Lambda(Y)}\right)^{2}.
\end{gather*}
Hence, rejecting concavity of $T(\cdot)$ (i.e. $T''(\cdot)<0$) implies that the baseline hazard is non-decreasing ($\lambda'(\cdot)\geq 0$). One can, thus, use the test of concavity of the transformation as a test for monotonicity of the baseline hazard, or in other words, as a test for positive duration dependence. In the economic context, one may be interested in detecting non-monotonicity of unemployment exit rate due to unemployment benefit exhaustion effects (see \cite{card_et_al07b} for discussion). 
\end{exmp}

Beyond these examples our procedure can also be used for specification search, i.e. determining if one should use a concave or convex transformation, and to test the curvature in wage regressions, e.g. if the effect of education or experience is concave, or the curvature of the marginal utility (or profit) function in hedonic models (see \cite{ekeland_et_al04}).%, if the assumption of selection on observables is met.

Our test statistic simply compares triples of $Y$'s corresponding to equally spaced index values $X'\beta_0$, thus it does not require estimation of the transformation function $T$ or the distribution of $\eps$. We only require an estimator of $\beta_0$ and, basically, the symmetry of the distribution of the error terms. Estimating $\beta_0$ can be done, for example, by using the maximum rank correlation estimator, \cite{han87}, or semiparametric least squares, \cite{ichimura93}. 

We propose both a global test that has power to detect globally convex or concave functions and leads to asymptotic normal critical values, as well as a more general test that detects local deviations from linearity, i.e. has power against alternatives that are both convex and concave on the domain of $T$. Our tests do not have a pivotal asymptotic distribution but we show that the critical values can be obtained by parametric bootstrap.

The cost of our test’s easy implementation is that the power properties of our local test are complex, and the test may not be consistent against some relevant alternatives. Nevertheless, our application demonstrates the usefulness of our testing approach in detecting nonconvexities in the demand for loans as a function of the interest rate.

Our statistic resembles the approach in \cite{abrevaya_jiang05} who test curvature in a nonparametric regression model. However, unlike their approach our test does not suffer from the curse of dimensionality due to the single index structure of the regression part and allows the marginal effect of covariates to vary with the error term (though in a manner restricted by the single-index). Also the details of the derivation of the asymptotic distribution are different due to presence of estimated $\beta_0$ in our statistic and somewhat distinct approach to obtaining power against general alternatives. These traits are shared by \cite{abrevaya_et_al10}, who test monotonicity in a generalised regression model with endogeneity, though a distinguishing feature of our work is that we formally show validity of bootstrap for obtaining   critical values.

Related literature includes tests for the sign of the treatment effect, see e.g. \cite{kline16}, and testing for treatment effect heterogeneity (\cite{abadie02}, \cite{crump_et_al08}, \cite{santanna21}, \cite{chernozhukov_et_al23}). Unlike the latter papers, our approach allows the treatment effect to vary with (independent) unobserved heterogeneity at the cost of imposing much more structure on the treatment effect model and the covariates. Testing curvature of the transformation can also be seen as a generalisation of specification testing in \cite{neumeyer_et_al16} and \cite{szydlowski20}. The idea of using curvature of the integrated hazard to test monotonicity of the baseline hazard has been utilised by \cite{hall_van_keilegom05}. Testing shape restrictions in a nonparametric regression model has been considered by \cite{ghosal_et_al00}, \cite{gutknecht16}, \cite{chetvertikov19} and \cite{komarova_hidalgo23}, among others. Similarly to this paper, \cite{chen_kato20} propose a bootstrap procedure for approximating the supremum of a local U-statistic.

The article is organized as follows. Section \ref{mainid} discusses the idea behind the testing procedure informally and the formal results are postponed till Sections \ref{formal}-\ref{sec:boot}. Sections \ref{MC}-\ref{appl} contain Monte Carlo results and our application to loan demand.  
Proofs, besides the proof of the main proposition, are located in the Appendix, which also contains some additional MC simulations.

\section{Main idea} \label{mainid}

\begin{figure}[ht] 
\caption{Testing concavity}
\begin{center}
\begin{tikzpicture}[scale=1.75]
\draw[->] (-2,-1)--(1.5,-1) node[right]{$Y$};
\draw[->] (-1,-2)--(-1,1.5) node[above]{$T(Y)$};
\draw[scale=1, domain=-1:1.5, smooth, variable=\x, blue, thick] plot ({\x}, {ln(\x+1.12)});
\node[black!60!green] at (-1,-0.5) [left] {$T(Y_{i})=X_{i}'\beta+\eps_{i} $};
\draw[dashed, black!60!green] (-1.05,-0.5) -- (-0.5134693,-0.5)--(-0.5134693,-1.05);
\node[black!60!green] at (-0.5134693,-1) [below] {$Y_{i}$};
\node[black!60!green] at (-1,0) [left] {$T(Y_{j})=X_{j}'\beta+\eps_{j} $};
\draw[dashed, black!60!green] (-1.05,0) -- (-0.12,0)--(-0.12,-1.05);
\node[black!60!green] at (-0.12,-1) [below] {$Y_{j}$};
\node[black!60!green] at (-1,0.5) [left] {$T(Y_{k})=X_{k}'\beta+\eps_{k} $};
\draw[dashed, black!60!green] (-1.05,0.5) -- (0.5287213,0.5)--(0.5287213,-1.05);
\node[black!60!green] at (0.5287213,-1) [below] {$Y_{k}$};
\end{tikzpicture}
\end{center}
\label{fig:intui}
%\floatfoot{Note: The arcs mark local parameter spaces corresponding to different $\theta_{2}\in \Theta_{2}(\hat{\overline{\theta}}_{1})$}  
\end{figure}
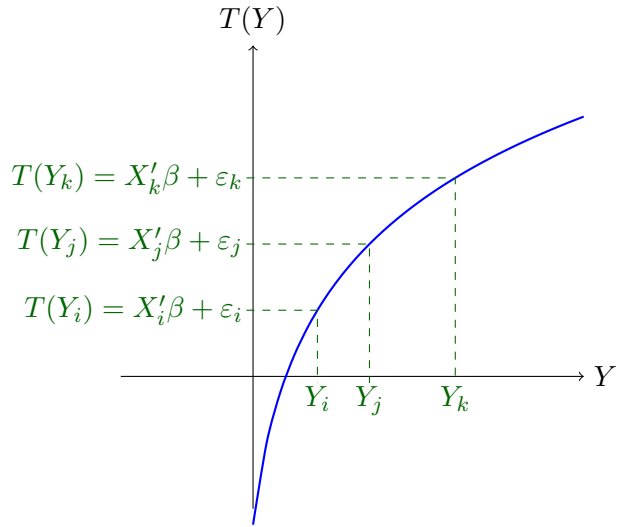

Figure \ref{fig:intui} portrays the intuition behind our test. The transformation plotted in the figure is concave and we display three ordered data points in the figure, $(Y_{i},Y_{j},Y_{k})$, for which the transformation function is equally spaced, i.e. $T(Y_{k})-T(Y_{j})=T(Y_{j})-T(Y_{i})$.

Concavity of $T(\cdot)$ implies that $Y_{k}-Y_{j}>Y_{j}-Y_{i}$. Note that $T(Y_{k})-T(Y_{j})=T(Y_{j})-T(Y_{i})$ is equivalent to $(X_{k}-X_{j})'\beta_0 +\eps_{k}-\eps_{j}= (X_{j}-X_{i})'\beta_0 +\eps_{j}-\eps_{i}$. Hence, ``on average'' equally spaced $T(Y)$'s mean equally spaced index values $X'\beta_0$. Therefore, we can detect deviations from concavity by considering the following criterion:
\begin{gather*} 
-\frac{1}{n(n-1)(n-2)}\sum_{i\neq j \neq k} \mathbbm{1}\{Y_{i}<Y_{j}<Y_{k} \} \mathbbm{1}\{Y_{k}-Y_{j}<Y_{j}-Y_{i} \}\mathbbm{1}\{X_{kj}'\beta_0 = X_{ji}'\beta_0\}
\end{gather*}
where $X_{ji} \equiv X_{j}-X_{i}$.

In order to make this criterion operational with continuous distribution of $X'\beta_0$ (which is required for identification) we need to replace the last indicator function with a smooth kernel $K_{h}(\cdot) = h^{-1}K(\cdot/h)$ and $\beta_0$ with its estimator $\hat{\beta}$:
\begin{gather*} 
-\frac{1}{n(n-1)(n-2)}\sum_{i\neq j \neq k} \mathbbm{1}\{Y_{i}<Y_{j}<Y_{k} \} \mathbbm{1}\{Y_{k}-Y_{j}<Y_{j}-Y_{i} \}K_{h}\left(X_{kj}'\hat{\beta} - X_{ji}'\hat{\beta}\right)
\end{gather*}
Deviations from convexity can be detected in a similar fashion.
Finally, we can combine both criterion functions to detect deviations from linearity:
\begin{gather} 
U_{n}=\frac{1}{n(n-1)(n-2)}\sum_{i\neq j \neq k} \mathbbm{1}\{Y_{i}<Y_{j}<Y_{k} \} sgn(Y_{k}-2Y_{j}+Y_{i})K_{h}\left(X_{kj}'\hat{\beta} - X_{ji}'\hat{\beta}\right)
\end{gather}
Here, very negative values of the objective function signify convexity and large positive values mark concavity. Outside testing, one may use the above measure itself to characterise an ``average'' curvature of function $T$, e.g. large positive values suggest that the function is predominantly concave.

Note that our test requires only one-dimensional kernel smoothing, thus it does not suffer from the curse of dimensionality. Also, unlike the specification test in \cite{szydlowski20} it does not require estimation of the transformation function, a computationally intense task itself.

\section{Formal definition and asymptotic theory}\label{formal}

\subsection{Global test}

As the probability limit of the criterion functions introduced in the previous section depends on the extent of concavity, but these functions are asymptotically centred at known values under linearity, we form our procedure as a test of:
\begin{align*}
H_0:
\begin{cases}
\text{$T(\cdot)$ is concave} \\
\text{$T(\cdot)$ is linear} \\
\text{$T(\cdot)$ is convex} \\
\end{cases}
 \qquad vs \qquad
H_{A}: 
\begin{cases}  
\text{$T(\cdot)$ is non-concave} \\
\text{$T(\cdot)$ is non-linear} \\
\text{$T(\cdot)$ is non-convex} \\
\end{cases}
\end{align*}
depending on the question of interest.  We will use the test statistic 
\begin{gather*}
S_{n}=\sqrt{n} U_{n} 
%\frac{\sum_{i\neq j \neq k} \mathbbm{1}\{Y_{i}<Y_{j}<Y_{k} \} sgn(Y_{k}-2Y_{j}+Y_{i})K_{h}\left((X_{kj}-X_{ji})'\hat{\beta}\right)}{\sum_{i\neq j \neq k} \mathbbm{1}\{Y_{i}<Y_{j}<Y_{k} \} K_{h}\left((X_{kj}-X_{ji})'\hat{\beta}\right)}
\end{gather*}
and reject the null hypothesis at level $\alpha$ if $S_{n}<c_{\alpha}, |S_{n}|>c_{1-\alpha/2}$ or $S_{n}>c_{1-\alpha}$ depending on $H_{0}$, respectively, where $c_{\alpha}$ denotes an $\alpha$ quantile from an appropriate asymptotic distribution. 

Define $\eps_{ji}=\eps_j - \eps_i$. Proposition \ref{prop:plim} justifies our testing strategy. We need the following symmetry assumption:

\begin{customassmpt}{SYM}\label{A:SYM}
Define $f_{\Delta \eps}(\cdot,\cdot)$ to be the joint distribution of $(\eps_{ji}, \eps_{kj})$. Assume that $\{(X_{i},Y_{i})\}_{i=1}^n$ are i.i.d and that $f_{\Delta \eps}(\eps^{\Delta}_1,\eps^{\Delta}_2)= f_{\Delta \eps}(\eps^{\Delta}_2,\eps^{\Delta}_1)$ for all $(\eps^{\Delta}_1,\eps^{\Delta}_2)$. 
\end{customassmpt}

\begin{proposition} \label{prop:plim}
Under Assumption \ref{A:SYM}:  $U_{n} \to^{p} \theta$ as $n\to \infty$, where:
%\begin{enumerate}[(i)]
%\item 
(i) $\theta\geq 0$ if $T(\cdot)$ is globally concave, 
%\item 
(ii) $\theta = 0$ if $T(\cdot)$ is globally linear, 
%\item 
(iii) $\theta\leq 0$ if $T(\cdot)$ is globally convex.
%\end{enumerate}
\end{proposition}

\begin{proof}
Let $f_{\xi\xi}(\cdot,\cdot)$ denote the joint distribution of $(X_{ji}'\beta_0, X_{kj}'\beta_0)$.  By
standard arguments:
\begin{gather*}
U_{n} \to^{p}  \quad \int_{-\infty}^{+\infty} E[\mathbbm{1}\{Y_{i}<Y_{j}<Y_{k}\}sgn(Y_{k}-2Y_{j}+Y_{i})|X_{kj}'\beta_0=X_{ji}'\beta_0=\xi] f_{\xi\xi}(\xi,\xi)d\xi 
\end{gather*}
and we can write:
\begin{align*}
E[\mathbbm{1} & \{Y_{i}<Y_{j}<Y_{k}\}  sgn(Y_{k}-2Y_{j}+Y_{i})|X_{kj}'\beta_0 = X_{ji}'\beta_0=\xi] = \\
&=  E[sgn(Y_{k}-2Y_{j}+Y_{i})|Y_{i}<Y_{j}<Y_{k}, X_{kj}'\beta_0=X_{ji}'\beta_0 =\xi]P(Y_{i}<Y_{j}<Y_{k}, X_{kj}'\beta_0=X_{ji}'\beta_0=\xi) \\
& \equiv \tilde{\theta} P(Y_{i}<Y_{j}<Y_{k}, X_{kj}'\beta_0=X_{ji}'\beta_0=\xi)
\end{align*}
Further: 
\begin{align*}
\tilde{\theta} &= P(Y_{k}-2Y_{j}+Y_{i}>0|Y_{i}<Y_{j}<Y_{k}, X_{kj}'\beta_0 =X_{ji}'\beta_0=\xi) \\
& \qquad \qquad \qquad \qquad  - P(Y_{k}-2Y_{j}+Y_{i}<0|Y_{i}<Y_{j}<Y_{k}, X_{kj}'\beta_0=X_{ji}'\beta_0=\xi) \\
& \equiv (1) - (2) 
\end{align*}
and the sign of the probability limit of $U_{n}$ is determined by the sign of $\tilde{\theta}$ (for all $\xi$). For simplicity let $\Xi$ denote the conditioning event in the probabilities above and note that this event is equivalent to $\{\eps_{ji}>-\xi, \eps_{kj}>-\xi, X_{kj}'\beta_0=X_{ji}'\beta_0=\xi\}$.  

(i) We will show that $(1)\geq (2)$ under concavity of $T$ (i.e. convexity of $T^{-1}$). 
Denote $a=X_{j}'\beta_0+\eps_{j}$, $b=X_{i}'\beta_0+\eps_{i}$ and $\Delta = \xi + \eps_{kj}$. Note that the conditioning event $Y_{i}<Y_{j}<Y_{k}$ implies that $\Delta>0, a-b>0$ by monotonicity of $T$. We can rewrite the event $Y_{k}-2Y_{j}+Y_{i}>0$ as:
\begin{gather*}
T^{-1}(a+\Delta)-T^{-1}(a)>T^{-1}(a)-T^{-1}(b)
\end{gather*}
Conditional on $\Xi(\xi)$ this event is implied by $\eps_{kj}>\eps_{ji}$. To see that observe that the latter event is implied by $\Delta>a-b$, which under convexity of $T^{-1}$ (i.e. concavity of $T$)  gives the desired result. 
Thus, we have:
\begin{gather*}
P(Y_{k}-2Y_{j}+Y_{i}>0|\Xi) \geq P(\eps_{kj}>\eps_{ji}|\Xi).
\end{gather*}
On the other hand, $Y_{k}-2Y_{j}+Y_{i}<0 \iff T^{-1}(a+\Delta)-T^{-1}(a)<T^{-1}(a)-T^{-1}(b)$, which under concavity implies $\eps_{kj}< \eps_{ji}$ as we need $\Delta<a-b$ for this event to occur and by definition $\Delta = a-b +\eps_{kj}-\eps_{ji}$. Hence:
\begin{gather*}
P(Y_{k}-2Y_{j}+Y_{i}<0|\Xi) \leq P(\eps_{kj}<\eps_{ji}|\Xi).
\end{gather*}
which implies $\tilde{\theta} \geq 2 P(\eps_{kj}>\eps_{ji}|\Xi) - 1$ and in order to show that $\tilde{\theta} \geq 0$ we need $P(\eps_{kj}<\eps_{ji}|\Xi)=0.5$ but that follows from symmetry of $f_{\Delta \eps}(\cdot,\cdot)$ along the 45$^{\circ}$ line.

(ii) It is enough to note that if $T$ is linear $P(Y_{k}-2Y_{j}+Y_{i}<0|\Xi) = P(\eps_{kj}<\eps_{ji}|\Xi)$, which implies $\tilde{\theta}=0$.

(iii) This part follows from an argument mirroring the one in (i).
\end{proof}

\begin{remark}
By direct calculation $f_{\Delta \eps}(\eps^{\Delta}_1,\eps^{\Delta}_2) = \int f(\eps - \eps^{\Delta}_1)f(\eps)f(\eps + \eps^{\Delta}_2) d\eps$. Therefore, $f_{\Delta \eps}(\eps^{\Delta}_1,\eps^{\Delta}_2)$ is symmetric along the 45$^{\circ}$ line if the probability density function of $\eps$, $f$, is symmetric around zero. 
Note that symmetry of the error term distribution is assumed in \cite{abrevaya_jiang05}. 
\end{remark}

\begin{remark}
If we defined our test statistic using two independent pairs of observations, namely:
\small
\begin{gather*}
\tilde{S}_{n}=\frac{\sqrt{n} }{n(n-1)(n-2)(n-3)}\sum_{i\neq j \neq k \neq l} \mathbbm{1}\{Y_{i}<Y_{j}<Y_{k}<Y_{m} \} sgn(Y_{m}-Y_{k}-Y_{j}+Y_{i})K_{h}\left((X_{mk}-X_{ji})'\hat{\beta}\right)
\end{gather*}
\normalsize 
the requirement for symmetry of $f_{\Delta \eps}(\eps^{\Delta}_1,\eps^{\Delta}_2)$ could potentially be dropped. This would, however, come at the increased computational cost as we have a 4-th order U statistic now. Also as we require four observations on $Y$ with approximately equally spaced index values $X'\beta_0$ instead of triples in our original statistic, the test based on $\tilde{S}_{n}$ is likely to have lower finite sample power than our baseline test.
\end{remark}

In order to obtain critical value for our tests we will assume that the model under the null hypothesis is linear. This is the worst-case $H_{0}$ for testing concavity/convexity as any small local deviation from linearity violates the hypothesis. This can also be seen from the proof of Proposition \ref{prop:plim}. In order to derive the asymptotic distribution of our statistic we make the following assumptions.

\begin{assumption}
\label{A1}
\begin{enumerate}[(a)]
\item The kernel $K(\cdot)$ is a bounded, nonnegative, symmetric, twice continuously differentiable function with support on $[-1,1]$ and uniformly bounded derivatives satisfying: \label{A1a}
\begin{enumerate}[(i)]
\item $\int K(s) ds = 1$,
\item $\int s^{2} K(s) ds < \infty$.
\item  \label{A1aiii} Let $\mathcal{K}$ be the antiderivative of $K$. We have: 
\begin{gather*}
\int \int K(s_1)K(s_2) [\mathcal{K} (2s_1-s_2) + 2 \mathcal{K} (2s_2-s_1)] s_1 d s_1 d s_2 > 0
\end{gather*}
\end{enumerate}
\item $h (\log n)^5 \to 0$ and 
$n h^3 \to \infty$ as $n\to \infty$. \label{A1b} 
\item Conditional on the remaining regressors, the distribution of the first element of $X$ is absolutely continuous with respect to the Lebesgue measure, with bounded and twice continuously differentiable density and uniformly bounded second derivatives. Each element of $X$ has a finite fourth moment. \label{A1c}
\item The density of $\eps$ is bounded and twice continuously differentiable, the derivatives are uniformly bounded. 
\item The estimator of $\beta_0$, $\hat{\beta}$, satisfies: \label{A1d}
\begin{gather*}
\hat{\beta}-\beta_0 = \frac{1}{n} \sum_{i=1}^{n}\Omega(X_{i}, Y_i) + o_{p}(n^{-1/2}).
\end{gather*}
where 
$E[\Omega(X_{i}, Y_i)]=0$ and each element of $\Omega$ has a finite fourth moment and is continuous in the second argument. 
\end{enumerate}  
\end{assumption}   

Assumption \ref{A1}\eqref{A1a}(iii) is nonstandard. It is not used for our global test, $S_n$, in this section but is needed for our local test in the next section to have power. Technically, the condition implies that the mean of our local statistic is negative (see proof of Theorem \ref{thm:bootpwr}). This assumption is satisfied by frequently used kernels like standard normal or Epanechnikov (with integral taking values 0.0072 and 0.0177, respectively). The bandwidth rate condition in Assumption \ref{A1}\eqref{A1b} is rather weak and allows standard ``rule-of-thumb'' bandwidth choice $h \sim n^{-1/5}$. Assumption \ref{A1}\eqref{A1d} requires $\hat{\beta}$ to be consistent %under our null hypothesis of linearity and 
under both the null and alternative hypothesis and is satisfied by a wide range of estimators including Han's MRC or Ichimura's semiparametric least squares (see e.g. appendix in \cite{szydlowski20} for discussion). 

Let $\xi_{ji}$ denote the index $X_{ji}'\beta_0$ and $f_{\xi|\xi}$ denote the density of $\xi_{kj}$ given $\xi_{ji}$. Define:
\begin{align*}
H(Y_{i},X_{i},\xi_{ji},\xi_{kj}) &= 2E[\mathbbm{1}\{Y_{i}<Y_{j}<Y_{k}\}sgn(Y_{k}-2Y_{j}+Y_{i})|Y_{i},X_{i},\xi_{ji},\xi_{kj}]\\
& \quad+2E[\mathbbm{1}\{Y_{j}<Y_{i}<Y_{k}\}sgn(Y_{k}-2Y_{i}+Y_{j})|Y_{i},X_{i},\xi_{ji},\xi_{kj}]\\
& \quad +2E[\mathbbm{1}\{Y_{j}<Y_{k}<Y_{i}\}sgn(Y_{i}-2Y_{k}+Y_{j})|Y_{i},X_{i},\xi_{ji},\xi_{kj}]\\
\delta(Y_{i},X_{i},\xi_{ji},\xi_{kj}) &= 3 H(Y_{i},X_{i},\xi_{ji},\xi_{kj})f_{\xi|\xi}(\xi_{kj}|\xi_{ji}=(X_{j}-X_{i})'\beta_0) \\
G(\xi_{ji},\xi_{kj}) &= 6E[\mathbbm{1}\{Y_{i}<Y_{j}<Y_{k}\}sgn(Y_{k}-2Y_{j}+Y_{i})(X_{kj}-X_{ji})'|\xi_{ji},\xi_{kj}]\\
\mu(\xi_{ji},\xi_{kj}) &=  4G(\xi_{ji},\xi_{kj})f_{\xi|\xi}(\xi_{kj}|\xi_{ji})
\end{align*}
and let $\mu_{2}$ denote the derivative of $\mu$ with respect to the second argument.

\begin{theorem}\label{thm:global}
If Assumption \ref{A1} holds, 
we have:
\begin{gather*}
\sqrt{n}(U_{n}-\theta) \to^{d} N(0,E[\psi_{i}^{2}])
\end{gather*}
where $\psi_{i}= E[\delta(Y_{i},X_{i},\xi_{ji},\xi_{ji})|Y_{i},X_{i}]- E[\mu_{2}(\xi_{ji},\xi_{ji})]\Omega(X_{i}, Y_i)$.
\end{theorem}

As linearity is the boundary case for testing $H_{0}$: $T(\cdot)$ is concave, or $H_{0}$: $T(\cdot)$ is convex, we will reject concavity if $S_{n}<c_{\alpha}$ and reject convexity if $S_{n}>c_{1-\alpha}$, where $c_{\alpha}$ denotes the $\alpha$ quantile from the normal asymptotic distribution. As typical with U-statistics, estimating the variance of the asymptotic distribution of $S_{n}$ is difficult. On the one hand, the plug-in estimator will involve estimating derivatives of conditional moments and distributions, which requires delicate choices of bandwidths and, essentially, calculation of higher order U-statistics. On the other hand, using the standardisation approach in \cite{ghosal_et_al00} will lead to a 5-th order U-statistic, which would be difficult to calculate with sample sizes typically encountered in applications.\footnote{Proceeding as there would involve calculating:
\scriptsize
\begin{align*}
\hat{\sigma}^{2} = \frac{1}{n(n-1)(n-2)(n-3)(n-4)}\sum_{i\neq j \neq k \neq l \neq m} &(\mathbbm{1}\{Y_{i}<Y_{j}<Y_{k}\}sgn(Y_{k}-2Y_{j}+Y_{i})K_{h}\left(X_{kj}'\hat{\beta} - X_{ji}'\hat{\beta}\right)+\tilde{\mu}_{2}\Omega_{i})\\
&\times (\mathbbm{1}\{Y_{i}<Y_{l}<Y_{m}\}sgn(Y_{m}-2Y_{l}+Y_{i})K_{h}\left(X_{ml}'\hat{\beta} - X_{li}'\hat{\beta}\right)+\tilde{\mu}_{2}\Omega_{i})\\
& + \text{symmetric terms},
\end{align*}
\footnotesize
where $\tilde{\mu}_{2}$ is an estimator of $-E[\mu_{2}(\xi_{ji},\xi_{ji})]$ (which itself is a 3-rd order U statistic).
}
Thus, in Section \ref{sec:boot} we resort to bootstrap for calculating the critical value as bootstrapping involves only repeated calculation of a 3-rd order U-statistic, $U_{n}$, which we find computationally easier than the aforementioned methods. 

The global test introduced in this section only has power against global deviations from concavity/linearity/convexity and does not have power if the function is both convex and concave on different parts of the domain. Thus, it can be used as a first check -- for example, if the test rejects concavity, the researcher concludes that the function cannot be globally concave. Failure to reject would, then, mean that one has to consider our local test described in the next section, in order to verify if indeed the function is globally concave.

\subsection{Local test}
The main idea of the local test is to consider only triples of the kind portrayed in Figure \ref{fig:intui} local to a point $y$ in the domain of the transformation function. In other words, we will check if the transformation function is concave/linear/convex locally around $y$. Local concavity may be of interest by itself. However, usually we are interested in verifying if the treatment effect is concave on the whole domain, thus we will take the minimum of the local statistics at different points $y$ to run the test. 

Formally, define:
\begin{align*}
U_{n}(y) = \frac{1}{n(n-1)(n-2)}\sum_{i\neq j \neq k} \mathbbm{1}\{Y_{i}<Y_{j}<Y_{k} \} & sgn(Y_{k}-2Y_{j}+Y_{i})  K_{h}(Y_{i}-y)K_{h}(Y_{j}-y)\\
& \quad \quad \quad \quad \times K_{h}(Y_{k}-y)K_{h}\left((X_{kj} - X_{ji})'\hat{\beta}\right)
\end{align*}
where in practice the bandwidth used for $(Y_{i}-y)$ can be different than for $(X_{kj} - X_{ji})'\hat{\beta}$ but, in order to simplify exposition and mathematical arguments, we assume that both bandwidths are of the same order and denote both by $h$. Now our local test statistic for testing concavity is defined as:
\begin{gather*}
S_{n}^{conc} = \inf_{y \in \mathcal{Y}} \sqrt{nh} U_{n}(y) 
\end{gather*}  
where $\mathcal{Y}$ is a compact set, and the statistic for testing convexity is defined with $\sup$ replacing $\inf$ above. Linearity can be tested by replacing $U_{n}(y)$ with its absolute value. For the rest of the article we concentrate on $S_{n}^{conc}$ as results for testing convexity and linearity follow by very similar arguments.

Intuitively, low values of $S_{n}^{conc}$ show that there is a large deviation from concavity at some point $y$ and, hence, the treatment effects are not accelerating on the whole domain of the outcome.\footnote{Note that by the formulas in Example 1 concavity of $T$ is equivalent to convexity of the outcome in the treatment.} Therefore, the null hypothesis of concavity would be rejected if $S_{n}^{conc} < c_{\alpha}^{conc}$ where $c_{\alpha}^{conc}$ is an appropriate quantile from the asymptotic approximation to the distribution of our statistic. 

In order to obtain $c_{\alpha}^{conc}$ one could imagine proceeding as in \cite{ghosal_et_al00}: approximate the standardised U-statistic process $\sqrt{n} U_{n}(y)/\sigma_{n}(y)$, where $\sigma_{n}(y)$ is the estimator of the asymptotic variance, by a Gaussian process and then apply the extreme value theory in order to derive the distribution of the infimum. However, pursuing that approach is difficult in our setup for two reasons: 1) under $H_0$ we still have $E[U_n(y)]=O(h)$ rather than $o(h)$ therein, 2) as discussed above, estimating $\sigma_{n}(y)$ is computationally expensive.
Instead we propose to use bootstrap to approximate $c_{\alpha}^{conc}$. 

Define:
\footnotesize
\begin{align*}
\phi_{i,n}(y) =& 6(E[ \mathbbm{1}\{Y_{i}<Y_{j}<Y_{k} \} sgn(Y_{k}-2Y_{j}+Y_{i})K_{h}(Y_{i}-y)K_{h}(Y_{j}-y)K_{h}(Y_{k}-y)K_{h}\left((X_{kj} - X_{ji})'\beta_0\right)|Y_{i},X_{i}]\\
&+E[ \mathbbm{1}\{Y_{j}<Y_{i}<Y_{k} \} sgn(Y_{k}-2Y_{i}+Y_{j})K_{h}(Y_{i}-y)K_{h}(Y_{j}-y)K_{h}(Y_{k}-y)K_{h}\left((X_{ki} - X_{ij})'\beta_0\right)|Y_{i},X_{i}]\\
&+E[ \mathbbm{1}\{Y_{j}<Y_{k}<Y_{i} \} sgn(Y_{i}-2Y_{k}+Y_{j})K_{h}(Y_{i}-y)K_{h}(Y_{j}-y)K_{h}(Y_{k}-y)K_{h}\left((X_{ik} - X_{kj})'\beta_0\right)|Y_{i},X_{i}]).
\end{align*}
\normalsize
The following asymptotic approximation will be useful in justifying our bootstrap procedure.

\begin{theorem}\label{thm:local}
If Assumption \ref{A1} holds, then:
\begin{gather*}
\sup_{y \in \mathcal{Y}}\left|U_{n}(y)  - E[U_n(y)] - \frac{1}{n} \sum_{i=1}^{n} \phi_{i,n}(y)\right| = o_{p}((nh)^{-1/2})
\end{gather*}
\end{theorem}  

An interesting consequence of this result is that 
the asymptotic distribution of the statistic does not depend on the estimation of $\beta_0$ as long as Assumption \ref{A1}\eqref{A1d} is satisfied,  unlike the global test (cf. Theorem \ref{thm:global}).

\section{Bootstrap critical values}\label{sec:boot}

Our bootstrap procedure for obtaining the critical value for the global or local test is as follows:
\begin{enumerate}
\item Estimate $\beta_0$ (e.g. by Han's MRC) and calculate the residuals $\hat{\eps}_{i} = Y_{i} - X_{i}'\hat{\beta}$.
\item For the global test: draw a random sample $\{v_i\}_{i=1}^{n}$ from a two point distribution with $P(v_{i}=-1)=P(v_{i}=1)=1/2$ and define $\eps_{i}^{*} =v_{i} \hat{\eps}_{i}$. For the local test: (a) draw $\{\eps_i^*\}_{i=1}^n$ with replacement from $\{\hat{\eps}_i\}_{i=1}^n$ or (b) follow the same procedure as for the global test. Then, generate $Y_{i}^*$ by: \label{boot2}
\begin{gather*}
Y_{i}^{*} = X_{i}'\hat{\beta} + \eps_i^*.
\end{gather*}
 \item Estimate $\beta_0$ (e.g. by Han's MRC) using the bootstrap sample. Let the resulting estimate be denoted by $\beta^*$.
 \item Calculate the statistic $S_{n}$ or $S_{n}^{conc}$ on the bootstrap sample using $\beta^*$ instead of $\hat{\beta}$. Denote the resulting bootstrap statistics by $S_{n}^{*}$ and $S_{n}^{conc,*}$.
 \item Obtain the empirical distribution of $S_{n}^*$ and $S_{n}^{conc,*}$ by repeating steps 1-4 many times. Calculate the $\alpha$ quantiles of the empirical distribution of $S_{n}^*$ and $S_{n}^{conc,*}$ and denote them by $c_{\alpha}^{*}$ and $c_{\alpha}^{conc,*}$, respectively.  
\end{enumerate} 

Step two imposes the null hypothesis of linearity on the bootstrap sample so it ``recentres'' the bootstrap statistic on the linear case. Furthermore, sampling from a symmetric two point distribution in the wild bootstrap imposes symmetry of the error distribution in the bootstrap sample, thus implying that Assumption \ref{A:SYM} is satisfied in that sample. 
Note that the jacknife multiplier bootstrap (JMB) of \cite{chen_kato20} will not work for the local test here (without modifications) as we have $E[U_n(y)]=O(h)$ even under the null so the JMB statistic will not be centred correctly.  
We assume an equivalent of Assumption \ref{A1}\eqref{A1d} for the bootstrap estimator $\beta^{*}$:
\begin{assumption}
\item The estimator of $\beta^*$ satisfies:
$\beta^* - \hat{\beta} = \frac{1}{n} \sum_{i=1}^{n}\Omega(X_{i}, Y_i^*) + o_{p}(n^{-1/2})$,
where 
$E[\Omega(X_{i}, Y_i^*)]=0$ has zero mean and each element of $\Omega$ has a finite fourth moment and is continuous in the second argument. 
\label{BA1d}
\end{assumption}
This assumption is satisfied in our leading case, when $\beta^*$ is estimated by MRC (see \cite{subbotin07}).

\begin{theorem}\label{thm:boot}
If Assumptions \ref{A1}-\ref{BA1d} hold
and $T(\cdot)$ is linear:
\begin{gather*}
\lim_{n\to \infty} P(S_{n}^{conc} \leq c_{\alpha}^{conc,*}) = \alpha
\end{gather*}
If, additionally, Assumption \ref{A:SYM} holds, we have:
\begin{gather*}
\lim_{n\to \infty} P\left(\sqrt{n}U_n \leq c_{\alpha}^{*}\right) = \alpha .
\end{gather*}
(and equivalent result holds for a test of linearity or convexity). 
\end{theorem}

Theorem \ref{thm:boot} implies, for example, that we can reject global concavity of $T$, i.e. increasing treatment effect, when $S_{n}^{conc}<c_{\alpha}^{conc,*}$, and reject global linearity when $|S_{n}^{conc}|>c_{1-\alpha/2}^{conc,*}$. We prove validity of bootstrap in unconditional probability, which is a weaker result than the standard result conditional on the sample (see e.g. \cite{cavaliere_georgiev20} for discussion), as the fact that $E[U_n(y)] = O(h)$ under $H_0$ and the need to use parametric bootstrap complicates the analysis. Note that we only require the symmetry of the error distribution for the global test. The next theorem characterises alternatives for which our tests are consistent. 

Assume that $c^{conc,*}_{\alpha}$ is generated using method (a) in step 2 above (see Appendix \ref{app:wildboot} for the analysis with the wild bootstrap in (b)). Let $f_{xb}$ denote the pdf of $X_i'\beta_0$ 
and $\mathbf{1} = [1 \quad  1 \quad  1]'$. Further, let $f_{\boldsymbol{\eps}}$ denote the joint distribution of $(\eps_i,\eps_j,\eps_k)$ and $\nabla f_{\boldsymbol{\eps}}$ denote its gradient and define $f_{\boldsymbol{\tilde{\eps}}}$,  $\nabla f_{\boldsymbol{\tilde{\eps}}}$, similarly for $\tilde{\eps} = Y -X'\beta_0$.

\begin{theorem}\label{thm:bootpwr}
If Assumptions \ref{A1}-\ref{BA1d} and Assumption \ref{A:SYM} hold,
and $T(\cdot)$ is globally convex:
\begin{gather*}
\lim_{n\to \infty} P\left(\sqrt{n}U_n \leq c_{\alpha}^{*}\right) = 1.
\end{gather*}
Moreover, if Assumptions \ref{A1}-\ref{BA1d} hold, $T(\cdot)$ is locally convex at some $y$ and the following condition holds: 
\scriptsize
\begin{align}
 &\sup_{y  \in \mathcal{Y}} \int \int \nabla f_{\boldsymbol{\tilde{\eps}}} \begin{pmatrix} y -\zeta +\xi \\ y -\zeta  \\ y -\zeta -\xi \end{pmatrix}' \mathbf{1}  f_{xb}(\zeta-\xi)f_{xb}(\zeta)f_{xb}(\zeta+\xi) d \zeta d\xi 
  \nonumber \\
& \quad - \sup_{y \in \mathcal{Y}} \int \int   \left\{ T'(y)^4 \nabla f_{\boldsymbol{\eps}}   \begin{pmatrix} T(y) -\zeta +\xi \\ T(y) -\zeta  \\ T(y) -\zeta -\xi  \end{pmatrix}  ' \mathbf{1} + 3T''(y)T'(y)^2 f_{\boldsymbol{\eps}} \begin{pmatrix} T(y) -\zeta +\xi \\ T(y) -\zeta   \\ T(y) -\zeta -\xi  \end{pmatrix} \right\} 
 f_{xb}(\zeta-\xi)f_{xb}(\zeta)f_{xb}(\zeta+\xi) d \zeta d\xi 
< 0,  \label{eq:pwrcond}
\end{align}
\normalsize
then we have:
\begin{gather*}
\lim_{n\to \infty} P(S_{n}^{conc} \leq c_{\alpha}^{conc,*}) = 1
\end{gather*}
(and equivalent result holds for a test of linearity or convexity). 
\end{theorem}

The condition in \eqref{eq:pwrcond} is technical and it is not straightforward to find general sufficient conditions for it to hold as it depends not only on the shape of $T$  but also the distributions of $\epsilon$ and $X$. 
However, we can make the following observations. 

\begin{remark}
For testing linearity (\emph{vel} testing treatment effect heterogeneity), the condition would only require the expression in \eqref{eq:pwrcond} to be nonzero and, thus, should be satisfied generically for non-linear $T(y)$'s, and the test should be consistent against (almost) all nonlinear alternatives. 
\end{remark}

\begin{remark}
Theorem \ref{thm:bootpwr} shows the cost of using a simple bootstrap test that avoids estimation of the transformation function, in particular the fact that our bootstrap does not correctly approximate the error distribution $f$ but rather distribution of OLS residuals, $\tilde{\eps}$. The latter distribution depends on the distribution of $X$ and on $T$, complicating the power properties of the test. In principle, one could try to estimate $T'$ and $\nabla f_{\boldsymbol{\eps}}$ and approximate the term:

\scriptsize
\begin{gather*}
 \sup_{y \in \mathcal{Y}} \int \int  T'(y)^4 \nabla f_{\boldsymbol{\eps}}   \begin{pmatrix} T(y) -\zeta +\xi \\ T(y) -\zeta  \\ T(y) -\zeta -\xi  \end{pmatrix}  ' \mathbf{1} f_{xb}(\zeta-\xi)f_{xb}(\zeta)f_{xb}(\zeta+\xi) d \zeta d\xi 
\end{gather*}
\normalsize
by a bootstrap procedure, making sure that condition \eqref{eq:pwrcond} would be satisfied.  However, this approach would substantially complicate the calculation of the critical value as estimators of $T$ and $f$ in the transformation model often involve non-smooth criterion functions and do not readily offer estimates of the derivatives (e.g. \cite{chen02}).
Thus, for practical reasons we prefer our current approach, even though it carries a cost in terms of power of the tests for concavity and convexity. 
\end{remark}

\begin{remark}
At an additional computational cost one could estimate $T$ and use semiparametric residuals: $\hat{\eps}_i = \hat{T}(Y_i) - X_i'\hat{\beta}$ in our bootstrap. We look at the latter case in Appendix \ref{app:MCnpar} where we show that the finite sample performance of such test does not dominate our local test. 
\end{remark} 

Finally, let us note that the computation of the local statistic involves evaluating a third order U-statistic at different points $y$ and, hence, takes significantly longer to compute than the global statistic. In practice, we recommend running the global test first and then proceed with the local test if the null hypothesis cannot be rejected in the first step.

\section{Monte Carlo simulations}\label{MC}
The data is generated from the following five models: 
\begin{align*}
Y &= X + \eps & \text{(D0)}\\
\log(Y + 2.12) - \log(2.12) &= X + \eps & \text{(D1)}\\
\frac{1}{13}\sinh(2Y) &= X + \eps & \text{(D2)}\\
\log(2.12) - \log(2.12-Y) &= X + \eps & \text{(D3)} \\
5 \tanh(0.5Y)& =  X + \eps & \text{(D4)} 
\end{align*}
where 
we draw $X$ and $\eps$ from the standard normal distribution. Note that D0 is the worst-case model in the null hypothesis, D1 imposes concavity of the transformation, D3 convexity and D2 and D4 are neither concave or convex. 

\begin{figure}[th] 
\centering
\caption{Monte Carlo designs}
\includegraphics[width=0.7\textwidth]{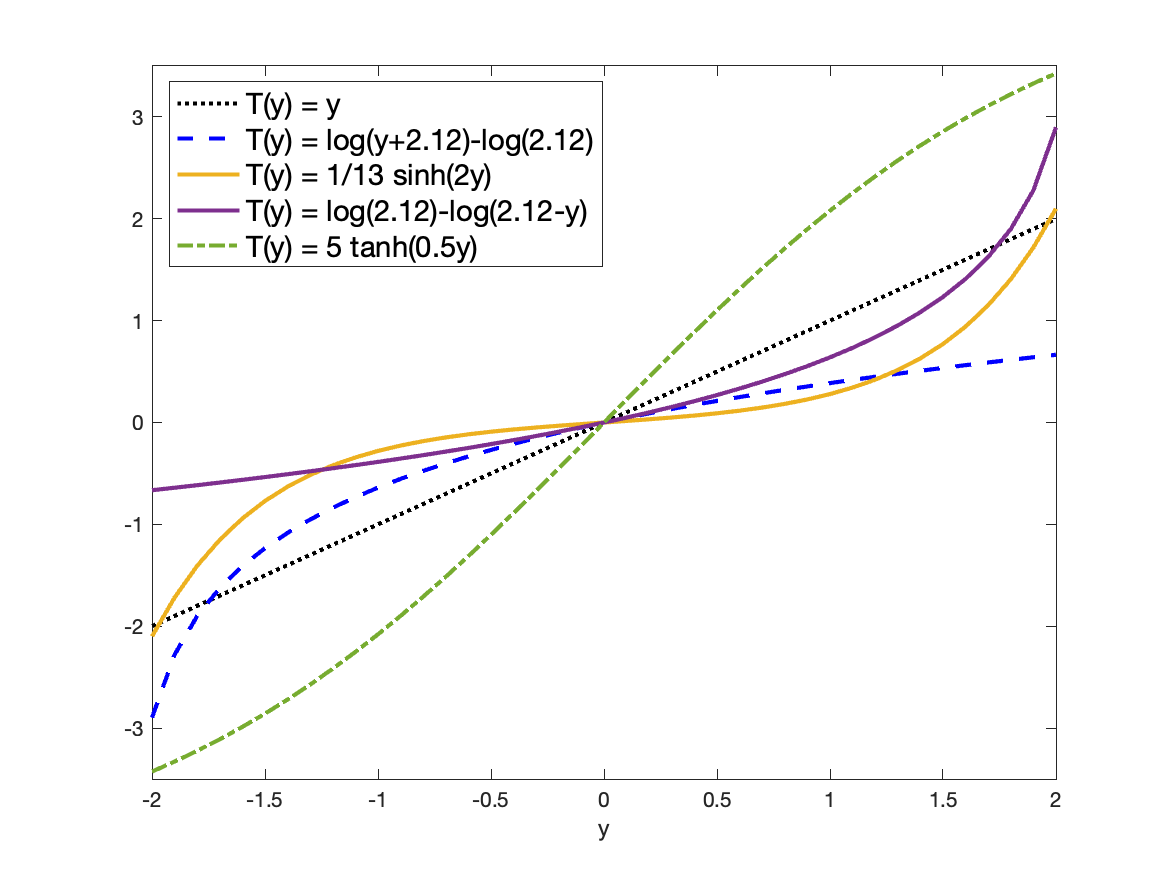}
\label{MCdesign1}
\end{figure}  

We run 1000 Monte Carlo replications. We use Gaussian kernel functions and rule-of-thumb bandwidths for both $(X_{kj}-X_{ji})'\hat{\beta}$ and $Y_{i}$, namely $h=1.06\hat{\sigma}n^{-1/5}$, 
where $\hat{\sigma}$ is a sample standard deviation of $(X_{kj}-X_{ji})'\hat{\beta}$ or $Y_{i}$.\footnote{The choice of bandwidth for $Y_i$ seems trickier as some transformations $T^{-1}$ make the distribution of $Y_i$ severly skewed. We have also tried using cross-validation (\emph{ucv} in R) and plug-in cross-validation (\emph{bcv} in R) with similar results.} 
In order to calculate the local statistic $S_{n}^{conc}$ we use a grid of values for $y$: -2:0.25:2, and take a minimum over the grid. The number of bootstrap replications used to calculate the critical value is 500 and we consider three sample sizes: $n=100, 250$ and $500$.

\begin{table}[ht]
	\centering
	\footnotesize
\begin{tabular}{ccccccccccc}			
\toprule																		
	&		&	\multicolumn{3}{c}{Global test}					&	\multicolumn{3}{c}{Local test}					&	\multicolumn{3}{c}{Local test - wild bootstrap}					\\
	&		&	$n=100$	&	$n=250$	&	$n=500$	&	$n=100$	&	$n=250$	&	$n=500$	&	$n=100$	&	$n=250$	&	$n=500$	 \\  \midrule
H0 true	&	D0	&	0.047	&	0.051	&	0.062	&	0.056	&	0.048	&	0.053	&	0.060	&	0.063	&	0.053	 \\
H0 true	&	D1	&	0.000	&	0.000	&	0.000	&	0.000	&	0.000	&	0.000	&	0.000	&	0.000	&	0.000	 \\
H0 false	&	D2	&	0.105	&	0.122	&	0.099	&	0.868	&	0.994	&	1.000	&	0.717	&	0.953	&	0.995	 \\
H0 false	&	D3	&	1.000	&	1.000	&	1.000	&	1.000	&	1.000	&	1.000	&	1.000	&	1.000	&	1.000	 \\
H0 false	&	D4	&	0.079	&	0.073	&	0.092	&	0.878	&	0.932	&	0.938	&	0.615	&	0.657	&	0.677	 \\
\bottomrule
\end{tabular}															
		\caption{Test of concavity, rejection probabilities, 5\% level}
	\label{tab:MC1}
		\floatfoot{Note: 1000 Monte Carlo simulations, 500 bootstrap replications.}
\end{table}
\normalsize

Table \ref{tab:MC1} contains the results of the Monte Carlo simulations. We concentrate on testing concavity, as the results for the linearity and convexity tests, are very similar. The rejection probabilities in the linear case (D0) are close to the nominal level for both the global and the local test and both tests have perfect power against a globally convex alternative in D3. 

As predicted, the global test has low power against D2 and D4, for which the transformation function is concave on part of the domain and convex on the other part. In this case deviations from concavity, as measured by our global statistic, cancel with positive values of the statistic obtained for the part of the domain where the function is concave, resulting in the global statistic taking values close to zero, just as for the linear case. 

For designs D2 and D4, our local test significantly improves over the global test with almost perfect detection of non-concavity in D2 for $n = 250$. The power against D4 increases slower with $n$ than D2 but we still reject H0 with close to certainty with $n=500$. This is in line with the intuition that the local test will concentrate on the region of the largest violation of concavity instead of averaging over the measures of concavity for different regions. 

Finally, note that, unlike for the global test, sampling from a symmetric distribution is not required in the local test bootstrap, so it is a question of finite sample performance if we would prefer the regular parametric bootstrap (part (a) in Step 2) or the symmetric wild parametric bootstrap in part (b). Comparing the last two panels in Table \ref{tab:MC1} shows that both bootstrap tests have the right level but the symmetric wild bootstrap test has lower power for our designs (see D2 and D4). 

\section{Application: Curvature of loan demand}\label{appl}

We use data from \cite{karlan_zinman08}, who ran randomised trials with a for-profit consumer lender in South Africa targeting high-risk consumer loan market. The lender randomised individual interest rate direct mail offers%to over 50,000 former clients
, conditional on the client’s risk category. \cite{karlan_zinman08} found that the loan demand curves are downward sloping. We will investigate if the demand curves are convex or, in other words, if increases in interest rate have a stronger negative effect on loan demand the lower the rate. 

\begin{figure}[h]
\centering
\caption{Estimated transformation}
\includegraphics[width=0.6\textwidth]{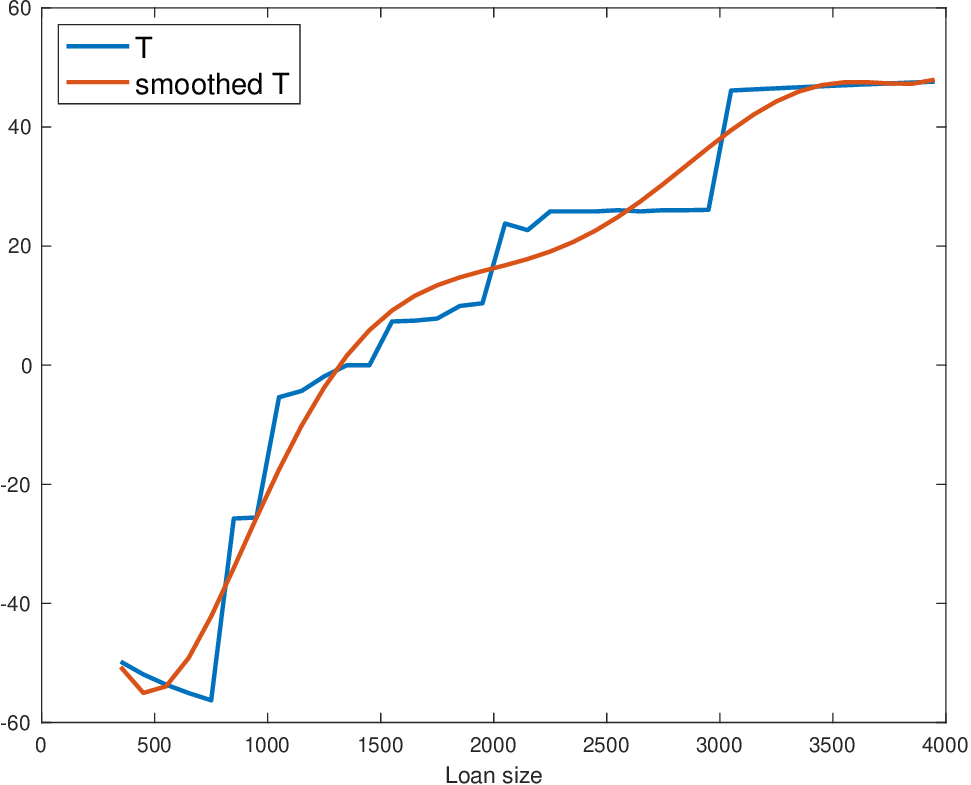}
\label{fig:Tappl}
\end{figure}  

The dependent variable is the amount borrowed (in rands) at the offered interest rate.\footnote{We abstract from selectivity issues here. See \cite{karlan_zinman08} for discussion.} The experiment was ran in three mailer waves over four months, thus as in \cite{karlan_zinman08} we include the risk category and wave dummies as controls in $X$. The data contains 2325 observations. As the first step, we estimated the transformation function in our model using the estimator in \cite{chen02} and plotted it in Figure \ref{fig:Tappl} together with a smoothed version.\footnote{This is for illustrative purposes only. Note that the estimator in \cite{chen02} does not impose monotonicity, so a full exercise of estimating $T$ should include additional step of monotonising the estimate. We want to stress that our testing procedure does not rely on any estimator of $T$.} The figure suggests that the transformation is not far from being concave on most of its domain, maybe besides the values of loan size below 1000 rands, which implies convex demand curve by formulas in Example 1.

\begin{table}[h]
	\centering
	\small
\begin{tabular}{c|ccc|cccccc}			
\toprule								
	&	\multicolumn{3}{c}{Global test}							&	\multicolumn{3}{|c}{Local test}							&	\multicolumn{3}{c}{Local test, $loan \geq 1000$} 					\\
	&				&	\multicolumn{2}{c|}{Reject $H_0$?}			&				&	\multicolumn{2}{c}{Reject $H_0$?}			&		&	\multicolumn{2}{c}{Reject $H_0$?}			\\
$H_0$	&	Statistic			&	5\%	&	10\%	&	Statistic			&	5\%	&	10\%	&	Statistic	&	5\%	&	10\%	 \\ \midrule 
convexity	&	0.26			&	No	&	No	&	$-0.27 \times 10^{-4}$			&	Yes	&	Yes	&	$0.19 \times 10^{-4}$	&	No	&	No	 \\
linearity	&	0.26			&	Yes	&	Yes	&				&		&		&		&		&		 \\
concavity	&	0.26			&	Yes	&	Yes	&				&		&		&		&		&		 \\
\bottomrule		
\end{tabular}
\caption{Testing curvature of loan demand}
\label{tab:appl}
\floatfoot{Note: For the local test we report the statistic multiplied by $n(n-1)(n-2)$. We choose bandwidth using cross-validation. 500 bootstrap replications. Data source: \cite{karlan_zinman08}, replication files.}											
\end{table}

In order to formally test our conjectures, we first apply our global test to verify if the demand function is concave, linear or convex.  As Table \ref{tab:appl} shows, global test rejects linearity and concavity of demand. Thus, in the second step we apply the local test to the null hypothesis of convexity and, in fact, reject the null hypothesis, concluding that the loan demand is not globally convex in the interest rate. 
To shed more light on where the non-convexity may come from, we re-ran the test on the sample excluding loans lower than 1000 rands and found that convexity is not rejected on this sub-sample. Therefore, overall we confirm that loan demand is mostly convex in interest rate, besides very small loans sector of the market.

\section{Conclusion}
Our application demonstrates usefulness of our testing procedures in recovering monotonicity of treatment effects. Particular appeal of the procedures described in this article comes from the fact that they avoid estimation of the transformation function and only require OLS estimation of the vector of coefficients $\beta_0$. Thus, they are relatively easy to implement. Additionally, computation of the third order U-statistic involved in our tests can be done efficiently by sorting the data by $Y$ first -- this reduces computational complexity to $O(nlog(n)+n(n-2)/2)$ from $O(n^{3})$ for a straightforward triple loop through the observations.    

\newpage
\section*{Appendix}
\appendix
\section{Proofs} \label{proofs}
Let
%\begin{gather*}
$\mathcal{G} = \{g_{y}(w_1,w_2,\ldots,w_m):  y \in \mathcal{Y} \subset \mathbb{R}\}$
%\end{gather*}
be a family of symmetric, real-valued functions defined on $\mathcal{W}^m$. We will use the operator notation common in the U-statistics literature. 
For example, for the case of $m=2$ we will have $P^0h = h$, $P^2h = \int \int g(w_1,w_2) dP(w_1) dP(w_2)$, $P_n g (w_1)=1/n \sum_{i=1}^n g(w_1,W_i)$  etc. We say that a symmetric function $g$ is $P$-canonical if $Pg(w_1,\ldots,w_{m-1},\cdot)=0$ for almost all $w_1,\ldots, w_{m-1}$.

Let $\|\cdot\|_{\mathcal{G}} \equiv \sup_{g \in \mathcal{G}}\|\cdot\|$ where $\|\cdot\|$ is the Euclidean norm and $\|\cdot\|_{\infty}, \|\cdot\|_{P,q}$ denote the sup and the $L_{q}(P)$ norm, respectively. Define a $U$-process:
\begin{gather*}
U_n^{(m)} g = \frac{(n-m)!}{n!} \sum_{\text{$i_1, i_2,\ldots,i_m$ distinct}} g(W_{i_1},W_{i_2},\ldots,W_{i_m})
\end{gather*}
Furthermore, define
\begin{gather*}
\pi_{k,m}^{P}g(w_1,\ldots,w_k) = (\delta_{w_1} - P)\ldots(\delta_{w_k}-P)P^{m-k} g 
\end{gather*}
where $\delta_{w_1}g =  g(w_1,\cdot)$.

Let us draw $\eps^*_i$ independently of $X_i$ and across observations. Thus, using $w=(x,\eps)$ we define $P_n^* g (w_1)=1/n \sum_{i=1}^n g(w_1,(X_i,\eps_i^*))$ and
\begin{gather*}
U_n^{*(m)} g = \frac{(n-m)!}{n!} \sum_{\text{$i_1, i_2,\ldots,i_m$ distinct}} g((X_{i_1}, \eps_{i_1}^*), (X_{i_2}, \eps_{i_2}^*), \ldots, (X_{i_m}, \eps_{i_m}^*))
\end{gather*}

We will sometimes use the following stochastic order arithmetic, for a sequence $a_n$:
\begin{gather*}
o^*_p(a_n) + o_p(a_n) = o_p(a_n), \quad  O^*_p(a_n) + O_p(a_n) = O_p(a_n)
\end{gather*}
which follows from the Law of Iterated Expectations. See \cite{szydlowski20} for more discussion. We will also write $\lesssim$ for inequality up to a multiplicative constant where the constant does not depend on the sample size $n$ or the sample data (but may depend on $m$ and characteristics of $\mathcal{G}$).

To economise on notation, throughout the proofs we write $\beta$ for $\beta_0$.
%Throughout the proofs $\beta$ denotes the probability limit of $\hat{\beta}$ both under the null hypothesis or under the alternative (under Assumption \ref{A1}\eqref{A1d}: $\beta=\beta_0$). We note here that Theorem \ref{thm:local} applies to any $\beta$ (not necessarily $\beta=\beta_0$) as long as Assumption \ref{A1}\eqref{A1d} is satisfied for that $\beta$ in place of $\beta_0$.

\subsection{Useful lemmas}
\begin{lemma} \label{lem:SZ}
\begin{enumerate}[(a)]
\item \label{lem:SZa} Let $\mathcal{G}$ be a class of P-canonical, Euclidean functions with envelope $G$. Then:
\begin{align*}
P&\|U_n^{(m)} g\|_{\mathcal{G}} \lesssim n^{-m/2}\sqrt{P^{m}G^{2}},\\
P&\|U_n^{*(m)} g\|_{\mathcal{G}} \lesssim n^{-m/2}\sqrt{P^{m}G^{2}}.
\end{align*}
\item \label{lem:SZb} Let $\mathcal{G}_n$ be a sequence of subclasses of $\mathcal{G}$ such that $\sup_{g \in \mathcal{G}_n} P^m g^2 \to 0$ as $n\to \infty$. Then:
\begin{gather*}
P\|U_n^{(m)} g\|_{\mathcal{G}_n} = o(n^{-m/2}).
\end{gather*}
\end{enumerate}
\end{lemma}

\begin{proof}

First part of (a) and (b) follow from Lemma 2 (a)-(b) in \cite{szydlowski20} with $p=1$, taking $h(\cdot)=g(\cdot)/\sqrt{P^{m}G^{2}}$ for the first result. 

The second part of (a) extends Lemma 2 (c) in \cite{szydlowski20} to parametric bootstrap. We give a short proof below for general $p$, putting emphasis on parts that do not follow \emph{verbatim} from the proof therein.

We have:
\begin{gather}
U_n^{*(m)}g = \sum_{k=0}^m {{m}\choose{k}} U_n^{*(k)}\left( \pi_{k,m}^{P_n}g\right) \label{bootU}
\end{gather}
Let $W_i = (X_i,\eps_i)$ and $W_i^* = (X_i,\eps_i^*)$. Since $\eps_i^*$'s are i.i.d., $\{W_i^*\}_{i=1}^n$ constitutes and i.i.d. sample, decoupling an repeated application of Jensen's inequality gives:
 \begin{gather*}
 P\left\| U_n^{*(k)}\left( \pi_{k,m}^{P_n}g \right) \right\|^p_{\mathcal{G}}  \lesssim P \left\| \frac{(n-k)!}{n!} \sum_{\text{$i_1,\ldots,i_{k}$ distinct}} \epsilon_{i_1}^{(1)}\ldots \epsilon_{i_k}^{(k)} P_n^{m-k} g(W^{*(1)}_{i_1},\ldots,W^{*(k)}_{i_k}) \right\|^p_{\mathcal{G}} \equiv (\star)
 \end{gather*}
where $\{W^{*(k)}_i: i=1,\ldots,n\}_{k=1}^m$ are i.i.d. copies of $\{W^{*}_i: i=1,\ldots,n\}$, $\{\epsilon_i:i=1,\ldots,n\}$ denote a sequence of Rademacher random variables independent of $W_i^*$'s and $\{\epsilon_i^{(k)}:i=1,\ldots,n\}_{k=1}^m$ are independent copies of $\{\epsilon_i:i=1,\ldots,n\}$.

Introduce 
$Z^*_{i_1} = \sum_{\text{$i_2,\ldots,i_{k} \neq i_1$ distinct}} \epsilon_{i_2}^{(2)}\ldots \epsilon_{i_k}^{(k)} P_n^{m-k} g(W^{*(1)}_{i_1},\ldots,W^{*(k)}_{i_k})$
and let $P_X$ and $P_{|X}$ denote expectation operators w.r.t. and conditional on $\{X_i\}_{i=1}^{\infty}$. Note that conditional on $X$  and
$\{\epsilon_{i_2}^{(2)},\ldots ,\epsilon_{i_k}^{(k)}, W^{*(2)}_{i_2},\ldots,W^{*(k)}_{i_k}\}$, $Z^*_{i_1}$'s are bootstrap draws from the sample $\{Z_i:i=1,\ldots,n\}$ where
$Z_{i_1} =  \sum_{\text{$i_2,\ldots,i_{k} \neq i_1$ distinct}} \epsilon_{i_2}^{(2)}\ldots \epsilon_{i_k}^{(k)} P_n^{m-k} g(W_{i_1},W^{*(2)}_{i_2},\ldots,W^{*(k)}_{i_k})$. Now using repeatedly Proposition 2.2 in \cite{gine_zinn90} (note that $Z^*_{i}$'s follow the empirical law of $Z_i$, cond. on $X$), Jensen's inequality and finally the result in Lemma 2(a) of \cite{szydlowski20} we obtain:
\small
\begin{align*}
(\star) &= P_X P_{|X} \left\| \frac{(n-k)!}{n!} \sum_{i_1} \epsilon_{i_1}^{(1)} Z^*_{i_1} \right\|^p_{\mathcal{G}} \lesssim  P_X P_{|X} \left\| \frac{(n-m)!}{n!} \sum_{\text{$i_1,\ldots,i_{m}$ distinct}} N_{i_1}^{(1)}\ldots N_{i_m}^{(m)} g(W_{i_1},\ldots,W_{i_m}) \right\|^p_{\mathcal{G}} \\
&\equiv P \left\| \frac{(n-m)!}{n!} \sum_{\text{$i_1,\ldots,i_{m}$ distinct}} \tilde{g}((W_{i_1}, N_{i_1}^{(1)}),\ldots, (W_{i_m}, N_{i_m}^{(1)})) \right\|^p_{\mathcal{G}} = O(n^{-pm/2})(P^{m}\tilde{G})^{p/2} \lesssim O(n^{-pm/2})(P^{m}G)^{p/2}
\end{align*}
\normalsize
where $N_i$'s are independent Poisson random variables with parameter 1/2. This and \eqref{bootU} imply the final result:
\begin{gather*}
\left(P\|U_n^{*(m)} g\|_{\mathcal{G}}^p\right)^{1/p} = O(n^{-m/2})\sqrt{P^{m}G^{2}}.
\end{gather*}
Now the result in the Lemma follows from setting $p=1$.
\end{proof}

\begin{lemma} \label{lem:Gh}
Let $\mathcal{G}$ be a Euclidean class of functions with envelope $G$. We have for $r\leq m$:
\begin{align*}
P\left(\left\|U_n^{(m)}g - \sum_{k=0}^{r-1} {{m}\choose{k}} U_n^{(k)}\pi_{k,m}^{P}g\right\|_{\mathcal{G}}\right) &\lesssim n^{-r/2} \sqrt{P^{m}G^{2}}
\end{align*}
\end{lemma}
\begin{proof}
From Theorem A.1 in \cite{ghosal_et_al00} and the discussion that follows we have:
\begin{gather*}
P\left(\left\|U_n^{(m)}g - \sum_{k=0}^{r-1} {{m}\choose{k}} U_n^{(k)}\pi_{k,m}^{P}g\right\|_{\mathcal{G}}\right) \lesssim n^{-r/2} \sqrt{P^{m}G^{2}} \int_{0}^{1} \sup_{Q: \text{$Q$ discrete}} \log N(\varepsilon \|G\|_{Q,2},\mathcal{G},L_{2}(Q))^{r/2}d\varepsilon
\end{gather*}
where $N(\cdot)$ denotes a covering number as in the Definition 2.1.5 in \cite{van_der_vaart_wellner96}. The integral on the right-hand side is bounded as $\mathcal{G}$ is Euclidean (ibid., Ch.2).
\end{proof}

\begin{lemma} \label{lem:Ch}
Assume that:
\begin{enumerate}[(a)]
\item The functional $B: \mathcal{G} \to \mathbb{R}$ satisfies: there exists a countable subset $\mathcal{\bar{G}}$ of $\mathcal{G}$ such that for any $g\in \mathcal{G}$, there exists a sequence $\bar{g}_{m} \in \mathcal{\bar{G}}$  with $\bar{g}_{m} \to g$ pointwise and $B(\bar{g}_{m})\to B(g)$. \label{lem:ChA}
\item The class of functions $\mathcal{G}$ is Euclidean with a measurable envelope $G$ and constants $A>0$ and $v \geq 1$. \label{lem:ChB}
\item There exist constants $b\geq \sigma >0$ and $q \in [4,\infty)$ such that $\sup_{g \in \mathcal{G}} P|g|^{k} \leq \sigma^{2}b^{k-2}$ for $k=2,3,4$ and $\|G\|_{P,q} \leq b$. \label{lem:ChC}
\end{enumerate}
Let $\mathbb{G}_{n}=\sqrt{n}P_{n}$
and $\mathbb{G}_{P}$ denote a centred Gaussian process indexed by $\mathcal{G}$ with covariance function $E[\mathbb{G}_{P}f \times \mathbb{G}_{P}g]=Cov(f(W),g(W))$. Let $N_{B}(\varepsilon)$ denote the bracketing number of the class of functions $\mathcal{BG}=\{B(g): g \in \mathcal{G}\}$. 

Then for every $\gamma \in (0,1)$ there exists a random variable $Z$ 
that follows  the same distribution as $\|B(g) +\mathbb{G}_{P}g\|_{\mathcal{G}}$ 
, such that:
\begin{align*}
|\|B(g) + \mathbb{G}_{n}g\|_{\mathcal{G}} - Z| &= O_{p}\left(\frac{bK_{n}}{\gamma^{1/q}n^{1/2-1/q}} + \frac{(b\sigma^{2}K_{n}^{2})^{1/3}}{\gamma^{1/3}n^{1/6}}\right) 
\end{align*}
where $K_{n}$ is of order $\log N_{B}(\varepsilon) + \log n \vee \log(Ab/\sigma)$ and $K_{n}^{3} \leq n$.
\end{lemma}

\begin{proof}
This result follows directly from Theorem 2.1 in \cite{chernozhukov_et_al16} with one caveat. \cite{chernozhukov_et_al16} require $\mathcal{G}$ to be VC-type, i.e. 
\begin{gather*}
\sup_Q N(\varepsilon \|G\|_{Q,2},\mathcal{G},L_{2}(Q)) \leq (A/\varepsilon)^v \quad \forall 0< \varepsilon \leq 1
\end{gather*}
where $Q$ belongs to the set of finitely discrete probability measures (see their Definition 2.1), with constants $A, v$ such that $A\geq e$ and $v\geq 1$. But we assume a Euclidean property, which following p. 789 in \cite{nolan_pollard87}, implies:
\begin{gather*}
\sup_Q N(\varepsilon \|G\|_{Q,2},\mathcal{G},L_{2}(Q)) \leq \tilde{A} \varepsilon^{-v} \quad \forall 0< \varepsilon \leq 1
\end{gather*}  
Thus, Euclidean property implies the VC-type property, taking $A=\tilde{A}^{1/v}$ and $\tilde{A}$ large enough..
\end{proof}

\begin{lemma}\label{lem:int}
Let $K$ be a symmetric kernel function supported on $[-1,1]$ and $\mathcal{K}$ be its antiderivative. We have:
\small
\begin{align*}
\int \int \int &\mathbbm{1}\{s_{1}<s_{2}<s_{3}\} sgn(s_{3}-2s_{2}+s_{1})K(s_{1})K(s_{2})K(s_{3}) d s_{1} d s_{2} d s_{3} =0\\
b_K \equiv \int \int \int &\Bigg\{(\mathbbm{1}\{s_{1}<s_{2}<s_{3}\} + \mathbbm{1}\{s_{3}<s_{2}<s_{1}\}) sgn(s_{3}-2s_{2}+s_{1}) + \\
&(\mathbbm{1}\{s_{1}<s_{3}<s_{2}\} + \mathbbm{1}\{s_{2}<s_{3}<s_{1}\}) sgn(s_{2}-2s_{3}+s_{1}) + \\
&(\mathbbm{1}\{s_{2}<s_{1}<s_{3}\} + \mathbbm{1}\{s_{3}<s_{1}<s_{2}\}) sgn(s_{3}-2s_{1}+s_{2})\Bigg\}
s_1 K(s_{1})K(s_{2})K(s_{3}) d s_{1} d s_{2} d s_{3}\\
&=  -2 \left[\int \int K(s_1)K(s_2) [\mathcal{K} (2s_1-s_2) + 2 \mathcal{K} (2s_2-s_1)] s_1 d s_1 d s_2 \right] \\
\int \int \int & (\mathbbm{1}\{s_{1}<s_{2}<s_{3}\} + \mathbbm{1}\{s_{3}<s_{2}<s_{1}\}) sgn(s_{3}-2s_{2}+s_{1}) 
s_k s_l K(s_{1})K(s_{2})K(s_{3}) d s_{1} d s_{2} d s_{3} \\
&= 0
\end{align*}
for any $k, l \in \{1,2,3\}$. 
\normalsize
\end{lemma}

\begin{proof}
For simplicity we write $\int_{-1}^{1}\int_{-1}^{1}\int_{-1}^{1}$ as $\int_{-1}^{1}$. We have:
\small
\begin{align*}
\int_{-1}^{1} \mathbbm{1}\{s_{1}<s_{2}<s_{3}\} & sgn(s_{3}-2s_{2}+s_{1})K(s_{1})K(s_{2})K(s_{3}) d s_{1} d s_{2} d s_{3} \\
 &= \int_{-1}^{1} \mathbbm{1}\{s_{1}<s_{2}<s_{3}\}  sgn(s_{3}-2s_{2}+s_{1})K(-s_{1})K(-s_{2})K(-s_{3}) d s_{1} d s_{2} d s_{3} \\
 &= \int_{-1}^{1}\mathbbm{1}\{\tilde{s}_{1}>\tilde{s}_{2}>\tilde{s}_{3}\}  sgn(-\tilde{s}_{3}+2\tilde{s}_{2}-\tilde{s}_{1})K(\tilde{s}_{1})K(\tilde{s}_{2})K(\tilde{s}_{3}) d \tilde{s}_{1} d \tilde{s}_{2} d \tilde{s}_{3}\\
 &= -\int_{-1}^{1}\mathbbm{1}\{\tilde{s}_{3}<\tilde{s}_{2}<\tilde{s}_{1}\}  sgn(\tilde{s}_{1}-2\tilde{s}_{2}+\tilde{s}_{3})K(\tilde{s}_{1})K(\tilde{s}_{2})K(\tilde{s}_{3}) d \tilde{s}_{1} d \tilde{s}_{2} d \tilde{s}_{3}
\end{align*}
\normalsize
where the first equality follows from symmetry of $K$, second from change of variables and the third from $sgn(-x)=-sgn(x)$. Finally note that the last integral is equal to the initial one after renaming the variables, so the result follows.

For the second equation in the lemma, note that the first part of the integral is the same as the second one if we rename $s_2$ to $s_3$. We have:
\small
\begin{align*}
\int &\int (\mathbbm{1}\{s_{1}<s_{2}<s_{3}\} + \mathbbm{1}\{s_{3}<s_{2}<s_{1}\}) sgn(s_{3}-2s_{2}+s_{1})
K(s_{2})K(s_{3}) d s_{2} d s_{3}\\
&= \int_{s_1}^1 \int_{s_2}^1 \mathbbm{1}\{s_3 - 2 s_2 +s_1>0\} K(s_{2})K(s_{3}) d s_{3} d s_{2} - \int_{s_1}^1 \int_{s_2}^1 \mathbbm{1}\{s_3 - 2 s_2 +s_1 < 0\} K(s_{2})K(s_{3}) d s_{3} d s_{2}  \\
& \quad +  \int_{-1}^{s_1} \int_{-1}^{s_2} \mathbbm{1}\{s_3 - 2 s_2 +s_1>0\} K(s_{2})K(s_{3}) d s_{3} d s_{2} - \int_{-1}^{s_1} \int_{-1}^{s_2} \mathbbm{1}\{s_3 - 2 s_2 +s_1 < 0\} K(s_{2})K(s_{3}) d s_{3} d s_{2}  \\
& = \int_{s_1}^1 \int_{2 s_2 - s_1}^1 K(s_{2})K(s_{3}) d s_{3} d s_{2} - \int_{s_1}^1 \int_{s_2}^{2 s_2 - s_1} K(s_{2})K(s_{3}) d s_{3} d s_{2} 
+  \int_{-1}^{s_1} \int_{2 s_2 - s_1}^{s_2} K(s_{2})K(s_{3}) d s_{3} d s_{2} \\
& \quad - \int_{-1}^{s_1} \int_{-1}^{2 s_2 - s_1}  K(s_{2})K(s_{3}) d s_{3} d s_{2} = \frac{3}{2} - \mathcal{K}(s_1) - 2 \int_{-1}^{1} K(s_2) \mathcal{K}(2 s_2-s_1) d s_2  
\end{align*}
\normalsize
and, similarly,  
\small
\begin{align*}
\int &\int (\mathbbm{1}\{s_{2}<s_{1}<s_{3}\} + \mathbbm{1}\{s_{3}<s_{1}<s_{2}\}) sgn(s_{3}-2s_{1}+s_{2})
K(s_{2})K(s_{3}) d s_{2} d s_{3}\\
& = \int_{-1}^{s_1} \int_{2 s_2 - s_1}^1 K(s_{2})K(s_{3}) d s_{3} d s_{2} - \int_{-1}^{s_1} \int_{s_1}^{2 s_1 - s_2} K(s_{2})K(s_{3}) d s_{3} d s_{2} 
+  \int_{s_1}^1 \int_{2 s_1 - s_2}^{s_1} K(s_{2})K(s_{3}) d s_{3} d s_{2} \\
& \quad - \int_{s_1}^1 \int_{-1}^{2 s_1 - s_2}  K(s_{2})K(s_{3}) d s_{3} d s_{2} = 2 \mathcal{K}(s_1) - 2 \int_{-1}^{1} K(s_2) \mathcal{K}(2 s_1-s_2) d s_2  
\end{align*}
\normalsize
which, using $\int K(s) s ds = 0$, gives the expression for the second integral in the lemma:
\small
\begin{align*}
\int & \int -\left[ 3 + 2  \int_{-1}^{1} K(s_2) \mathcal{K}(2 s_1-s_2) d s_2 + 4 \int_{-1}^{1} K(s_2) \mathcal{K}(2 s_2-s_1) d s_2  \right]
s_1 K(s_{1})K(s_{2})K(s_{3}) d s_{1} d s_{2} d s_{3} \\
& = -2 \int \int K(s_1)K(s_2) [\mathcal{K} (2s_1-s_2) + 2 \mathcal{K} (2s_2-s_1)] s_1 d s_1 d s_2.
\end{align*} 
The last equality in the lemma follows from symmetry of $K$ by similar argument as the one above, using change of variables.

\end{proof}

\begin{lemma}\label{lem:Eucl}
Let $K$ satisfy Assumption \ref{A1}(a). The classes of functions:
\scriptsize
\begin{align*}
\mathcal{F}_{1} =& \Bigg\{\mathbbm{1}\{x_{1}'\beta + \eps_1<x_{2}'\beta + \eps_2<x_{3}'\beta + \eps_3\}sgn(x_{3}'\beta +\eps_3 -2(x_{2}'\beta + \eps_2)+x_{1}'\beta + \eps_1)\\
& \hfill \times K\left(\frac{x_{1}'\beta + \eps_1-y}{h}\right)K\left(\frac{x_{2}'\beta + \eps_2-y}{h}\right)K\left(\frac{x_{3}'\beta + \eps_3-y}{h}\right)K\left(\frac{(x_{32}-x_{21})'\beta}{h}\right): y \in \mathcal{Y}, \beta \in \Theta_{\beta} \Bigg\} \\
\mathcal{F}_{2} =& \Bigg\{\mathbbm{1}\{x_{1}'\beta + \eps_1<x_{2}'\beta + \eps_2<x_{3}'\beta + \eps_3\}sgn(x_{3}'\beta +\eps_3 -2(x_{2}'\beta + \eps_2)+x_{1}'\beta + \eps_1)\\
& \hfill \times K\left(\frac{x_{1}'\beta + \eps_1-y}{h}\right)K\left(\frac{x_{2}'\beta + \eps_2-y}{h}\right)K\left(\frac{x_{3}'\beta + \eps_3-y}{h}\right)K'\left(\frac{(x_{32}-x_{21})'\beta}{h}\right) (x_{32}-x_{21})': y \in\mathcal{Y}, \beta \in \Theta_{\beta} \Bigg\} \\
\mathcal{F}_{3} =& \Bigg\{\mathbbm{1}\{x_{1}'\beta + \eps_1<x_{2}'\beta + \eps_2<x_{3}'\beta + \eps_3\}sgn(x_{3}'\beta +\eps_3 -2(x_{2}'\beta + \eps_2)+x_{1}'\beta + \eps_1)\\
& \hfill  \times K\left(\frac{x_{1}'\beta + \eps_1-y}{h}\right)K\left(\frac{x_{2}'\beta + \eps_2-y}{h}\right)K\left(\frac{x_{3}'\beta + \eps_3-y}{h}\right)K''\left(\frac{(x_{32}-x_{21})'\beta}{h}\right)  (x_{32}-x_{21})(x_{32}-x_{21})': y \in\mathcal{Y}, \beta \in \Theta_{\beta}\Bigg\} 
\end{align*}
\normalsize
are Euclidean with envelopes: 
\scriptsize
\begin{align*}
F_{1} &= \|K\|_{\infty}^{3}\mathbbm{1}\{|y_{1}-y_{2}|<2h\}\mathbbm{1}\{|y_{1}-y_{3}|<2h\}\mathbbm{1}\{|y_{2}-y_{3}|<2h\}K((x_{32}-x_{21})'\beta /h),\\
F_{2} &= \|K\|_{\infty}^{3}\mathbbm{1}\{|y_{1}-y_{2}|<2h\}\mathbbm{1}\{|y_{1}-y_{3}|<2h\}\mathbbm{1}\{|y_{2}-y_{3}|<2h\}K'((x_{32}-x_{21})'\beta /h)(x_{32}-x_{21})'\\
F_{3} &= \|K\|_{\infty}^{3}\mathbbm{1}\{|y_{1}-y_{2}|<2h\}\mathbbm{1}\{|y_{1}-y_{3}|<2h\}\mathbbm{1}\{|y_{2}-y_{3}|<2h\}K''((x_{32}-x_{21})'\beta /h)(x_{32}-x_{21})(x_{32}-x_{21})'
\end{align*}
\normalsize
\end{lemma}

\begin{proof}
Note that $\mathcal{F}_{1}$, $\mathcal{F}_{2}$ and $\mathcal{F}_{3}$ are products of functions $\mathbbm{1}\{\boldsymbol{\cdot}'\beta +\boldsymbol{\cdot}>y\}$ and classes of functions of the form $ K\left(\frac{\boldsymbol{\cdot}'\beta+\boldsymbol{\cdot}-y}{h}\right)$, $ K'\left(\frac{\boldsymbol{\cdot}'\beta+\boldsymbol{\cdot}-y}{h}\right)$ or $ K''\left(\frac{\boldsymbol{\cdot}'\beta+\boldsymbol{\cdot}-y}{h}\right)$. The former class is Euclidean by Lemma 2.4 in \cite{pakes_pollard89} and the latter by Lemma 22 in \cite{nolan_pollard87}. As the product of Euclidean classes is Euclidean, the result follows. The form of the envelope follows from the fact that $K$ is supported on $[-1,1]$. 
\end{proof}

\section{Proof of Theorem \ref{thm:global}}
To simplify notation set $\theta=0$. The argument for $\theta \neq 0$ follows \emph{verbatim} by recentering the test statistic appropriately. Write $U_{n}$ as:
\footnotesize
\begin{align*}
U_{n} =& \frac{1}{n(n-1)(n-2)}\bigg\{\sum_{i\neq j \neq k} \mathbbm{1}\{Y_{i}<Y_{j}<Y_{k} \} sgn(Y_{k}-2Y_{j}+Y_{i})K_{h}\left((X_{kj} - X_{ji})'\beta\right) &\text{(I)}\\
& + \sum_{i\neq j \neq k} \mathbbm{1}\{Y_{i}<Y_{j}<Y_{k} \} sgn(Y_{k}-2Y_{j}+Y_{i})K'_{h}\left((X_{kj} - X_{ji})'\beta\right)(X_{kj} - X_{ji})'(\hat{\beta}-\beta) &\text{(II)} \\
& + \sum_{i\neq j \neq k} \mathbbm{1}\{Y_{i}<Y_{j}<Y_{k} \} sgn(Y_{k}-2Y_{j}+Y_{i})K''_{h}\left((X_{kj} - X_{ji})'\tilde{\beta}\right)(\hat{\beta}-\beta)'(X_{kj} - X_{ji})(X_{kj} - X_{ji})'(\hat{\beta}-\beta)\bigg\} &\text{(III)} 
\end{align*}
\normalsize
where $\tilde{\beta}$ is between $\beta$ and $\hat{\beta}$.

Note that by Cauchy-Schwartz inequality the third term in this decomposition can be bounded as follows:
\footnotesize
\begin{gather*}
|(III)| \leq h^{-3}\|K''\|_{\infty} \|\hat{\beta}-\beta\|^{2} \left\| \frac{1}{n(n-1)(n-2)}\sum_{i\neq j \neq k} \mathbbm{1}\{Y_{i}<Y_{j}<Y_{k} \} sgn(Y_{k}-2Y_{j}+Y_{i})(X_{kj} - X_{ji})(X_{kj} - X_{ji})'\right\|
\end{gather*} 
\normalsize
Now $\|\hat{\beta}-\beta\|^{2} = O_{p}(n^{-1})$ by Assumption \ref{A1}\eqref{A1d} and the U-statistic on the right-hand side is $O_{p}(n^{-1/2})$ by Theorem 5.5.1A in \cite{serfling80}, which gives:
\begin{gather*}
(III) = O_{p}(n^{-3/2}h^{-3}) = o_{p}(n^{-1/2})
\end{gather*}
under our bandwidth conditions.

Further, using Assumption \ref{A1}\eqref{A1d} I can rewrite $(II)$ as:
\small
\begin{align*}
\frac{1}{n(n-1)(n-2)} &\sum_{i\neq j \neq k} \mathbbm{1}\{Y_{i}<Y_{j}<Y_{k} \} sgn(Y_{k}-2Y_{j}+Y_{i})K'_{h}\left((X_{kj} - X_{ji})'\beta\right)(X_{kj} - X_{ji})'P_{n}\Omega = \\
&= \frac{1}{n(n-1)(n-2)(n-3)} \sum_{i\neq j \neq k \neq l} \mathbbm{1}\{Y_{i}<Y_{j}<Y_{k} \} sgn(Y_{k}-2Y_{j}+Y_{i})K'_{h}\left((X_{kj} - X_{ji})'\beta\right)\\
& \quad \times (X_{kj} - X_{ji})'\Omega(X_l,Y_l) + o_{p}(n^{-1/2}).
\end{align*}
\normalsize

It remains to apply Lemma \ref{lem:Gh} to (I) and (II) with $r=2$ and $\mathcal{G}$ containing a single function. In order to do that, let $g^{I}$ and $g^{II}$ be symmetrised kernels of the U-statistics in $(I)$ and $(II)$. Note that $(I) = U_{n}^{(3)}g^{I}$ and $(II) = U_{n}^{4}g^{II} + o_p(n^{-1/2})$ and the envelopes satisfy:
\small
\begin{align*}
E(G^{I})^{2} &\lesssim h^{-2}E[K^2((X_{32}-X_{21})'\beta)/h)] = O(h^{-1})\\
E(G^{II})^2 &\lesssim h^{-4}E[(K'((X_{32}-X_{21})'\beta)/h))^2(X_{32}-X_{21})'\Omega_4] \\
& \leq  h^{-4}E[(K'((X_{32}-X_{21})'\beta)/h))^2(X_{32}-X_{21})'(X_{32}-X_{21})]E[\Omega_4'\Omega_4]   = O(h^{-3})
\end{align*}
\normalsize
which, using Lemma \ref{lem:Gh} gives:
\begin{gather*}
P\left(\left\|U_n^{(m)}(g^{I}+g^{II}) - 3P_n\pi_{1,3}^{P}g^{I} - 4P_n\pi_{1,4}^{P} g^{II}\right\|\right) = O(n^{-1}h^{-1/2} + n^{-1}h^{-3/2}) = o_{p}(n^{-1/2})
\end{gather*}
under our bandwidth conditions in Assumption \ref{A1}\eqref{A1b}.

Now by direct calculation, using change of variables, Taylor expansion and $\int K'(u) du = 0$, $\int K'(u) u du =-1$ we get $3\pi_{1,3}^{P}g^{I} = E[\delta(Y_{i},X_{i},\xi_{ji},\xi_{ji})|Y_{i},X_{i}]$, $4\pi_{1,4}^{P} g^{II} = - E[\mu_{2}(\xi_{ji},\xi_{ji})]\Omega(X_{i}, Y_i)$ and the result follows from the CLT.

\section{Proof of Theorem \ref{thm:local}}
Let $K_{h}(\mathbf{Y}-y)\equiv K_{h}(Y_{i}-y)K_{h}(Y_{j}-y)K_{h}(Y_{k}-y)$. Write $U_{n}(y)$ as:
\footnotesize
\begin{align}
& \frac{1}{n(n-1)(n-2)}\bigg\{\sum_{i\neq j \neq k} \mathbbm{1}\{Y_{i}<Y_{j}<Y_{k} \} sgn(Y_{k}-2Y_{j}+Y_{i})K_{h}(\mathbf{Y}-y)K_{h}\left((X_{kj} - X_{ji})'\beta\right) &\text{(I)} \nonumber\\ 
& + \sum_{i\neq j \neq k} \mathbbm{1}\{Y_{i}<Y_{j}<Y_{k} \} sgn(Y_{k}-2Y_{j}+Y_{i})K_{h}(\mathbf{Y}-y)K'_{h}\left((X_{kj} - X_{ji})'\beta\right)(X_{kj} - X_{ji})'(\hat{\beta}-\beta) &\text{(II)} \nonumber\\ 
& + (\hat{\beta}-\beta)'\sum_{i\neq j \neq k} \mathbbm{1}\{Y_{i}<Y_{j}<Y_{k} \} sgn(Y_{k}-2Y_{j}+Y_{i})K_{h}(\mathbf{Y}-y)K''_{h}\left((X_{kj} - X_{ji})'\tilde{\beta}\right)(X_{kj} - X_{ji})(X_{kj} - X_{ji})'(\hat{\beta}-\beta)\bigg\} &\text{(III)} \label{decomp}
\end{align}
\normalsize
where $\tilde{\beta}$ is between $\beta$ and $\hat{\beta}$.

First, we will show that $(III) = o_{p}(n^{-1/2})$ using Lemmas \ref{lem:SZ}-\ref{lem:Gh}. Define $g_{y}^{III}((X_i,Y_i),(X_j,Y_j),(X_k,Y_k))$ to be the symmetrised version of $\mathbbm{1}\{Y_{i}<Y_{j}<Y_{k} \} sgn(Y_{k}-2Y_{j}+Y_{i})K_{h}(\mathbf{Y}-y)K''_{h}\left((X_{kj} - X_{ji})'\tilde{\beta}\right)(X_{kj} - X_{ji})(X_{kj} - X_{ji})'$. Its envelope is given by:
\small
\begin{align*}
G^{III} &= h^{-6} \|K\|^3_{\infty}\mathbbm{1}\{|y_1-y_2|<2h\} \mathbbm{1}\{|y_1-y_3|<2h\} \mathbbm{1}\{|y_2-y_3|<2h\}K''((x_{32}-x_{21})'\tilde{\beta}/h) \\
& \quad \qquad \qquad \hspace{10cm} \times (x_{32}-x_{21}) (x_{32}-x_{21})'   
\end{align*}
\normalsize

Let $f_{Y|X}$ denote the joint distribution of $(Y_1,Y_2,Y_3)$ given $(X_1,X_2,X_3)$. By standard calculation and Lemma \ref{lem:int}:
\small
\begin{align}
P^3 g_{y}^{III} & = hE\bigg[\left(\int \mathbbm{1}\{s_{1}<s_{2}<s_{3}\} sgn(s_{3}-2s_{2}+s_{1})K(\mathbf{s})\mathbf{s}' 
\nabla f_{Y|X}(\mathbf{y}|X) d\mathbf{s} \right) K''_{h}\left((X_{kj} - X_{ji})'\tilde{\beta}\right) \nonumber \\ 
& \quad \times (X_{kj} - X_{ji})(X_{kj} - X_{ji})' \bigg] + o(h) = O(h) \label{eq:EIII}
\end{align}
\normalsize

 Note that Lemma \ref{lem:Eucl} applies to the class of functions $\{g_{y}^{III}: y \in \mathcal{Y}\}$, thus this class is Euclidean. Further note that expectation of the square of the envelope of the $(l,m)$-th element of  $g_{y}^{III}$ is:
 \small
 \begin{align*}
 P^3 (G^{III}_{lm})^2 &\cong h^{-12} E[P(|Y_1-Y_2|<2h, |Y_1-Y_3|<2h, |Y_2-Y_3|<2h|X)K''((x_{32}-x_{21})'\tilde{\beta}/h)(X_{l,32}-X_{l,21})^2 \\
 & \quad \times (X_{m,32}-X_{m,21})^2] = O(h^{-9})  
 \end{align*}
 \normalsize
 using dominated convergence theorem, as $P(|Y_1-Y_2|<2h, |Y_1-Y_3|<2h, |Y_2-Y_3|<2h|X) = O(h^2)$ 
 and each element of $X$ has a finite fourth moment. Now Lemma \ref{lem:Gh} gives:
 \begin{gather}
 U_{n}^{(3)}g_{y}^{III} = P^3 g_{y}^{III} + 3 P_n \pi_{1,3}^P g_{y}^{III} +  O_p(n^{-1}h^{-9/2}). \label{eq:IIIdcmp}
 \end{gather}
 and by Lemma \ref{lem:SZ}:
 \begin{gather}
 P_n \pi_{1,3}^P g_{y}^{III} = O_p(n^{-1/2} h^{-7/2}) \label{eq:IIIfoterm}
  \end{gather}
  using the fact that the envelope of $\pi_{1,3}^P g_{y}^{III}$ satisfies: $P(\pi_{1,3}^P G^{III})^2 \cong h^{-8} P(|Y_2-Y_3|<2h) \cong h^{-7}$.
  
Now using \eqref{eq:EIII}, \eqref{eq:IIIdcmp}, \eqref{eq:IIIfoterm} and Assumption \ref{A1}\eqref{A1d} we can bound $(III)$ uniformly over $y$:
\begin{gather}
(III) \lesssim \| \hat{\beta}-\beta\|^2 ( O_{p}(h) + O_{p}(n^{-1/2}h^{-7/2}) + O_p(n^{-1}h^{-9/2})) = O_{p}(n^{-1}h + n^{-3/2}h^{-7/2} + n^{-2}h^{-9/2}) \label{eq:IIIfinal}
\end{gather} 
and this is $o_{p}((nh)^{-1/2})$ under our rate conditions. 

Further, let $g_{y}^{II}$ denote the symmetrised version of $\mathbbm{1}\{Y_{i}<Y_{j}<Y_{k} \} sgn(Y_{k}-2Y_{j}+Y_{i})K_{h}(\mathbf{Y}-y)K'_{h}\left((X_{kj} - X_{ji})'\beta\right)(X_{kj} - X_{ji})'\Omega(X_{l}, Y_l)$. Note that we have $(II) = U_{n}^{(4)}g_{y}^{II} + o_{p}(n^{-1/2})$. The envelope satisfies:
\begin{align*}
P^4(G^{II})^2 &\cong h^{-10} E[P(|Y_1-Y_2|<2h, |Y_1-Y_3|<2h, |Y_2-Y_3|<2h|X) K'((X_{32}-X_{21})'\beta /h) \\
 & \quad \hspace{4cm} \times ((X_{32}-X_{21})'\Omega_4)^2] = O(h^{-7}),
\end{align*}
where we have used the fact that $X$ and $\Omega$ have finite fourth moments, which by Lemma \ref{lem:Gh} 
implies:
 \begin{gather*}
 U_{n}^{(4)}g_{y}^{II} = P^4 g_{y}^{II} + 4P_n \pi_{1,4}^P g_{y}^{II} + {{4} \choose {2}} U_n^{(2)} \pi_{2,4}^P g_{y}^{II} + {{4} \choose {3}} U_n^{(3)} \pi_{3,4}^P g_{y}^{II} +  O_p(n^{-2}h^{-7/2}). \label{eq:IIdcmp}
 \end{gather*}
Now $P^4 g_{y}^{II}=0$ since $P\Omega=0$. Further, the envelopes of the second, the third and the fourth term above satisfy: $P(\pi_{1,4}G^{II})^2 = O(h^{2})$, $P(\pi_{2,4}G^{II})^2 = O(h^{-2})$ and $P(\pi_{3,4}G^{II})^2 = O(h^{-3})$  (where we used $\int K'(u) du =0$ and boundedness of the moments of $\Omega$ and $X$), which by Lemma \ref{lem:SZ} implies:
\begin{gather}
 U_{n}^{(4)}g_{y}^{II} = O_p(n^{-1/2}h + n^{-1}h^{-1} + n^{-3/2} h^{-3/2} +n^{-2}h^{-7/2})  = o_p((nh)^{-1/2}) \label{eq:IIfinal}
\end{gather}
 where the last equality follows from Assumption \ref{A1}\eqref{A1b}.

Additionally, if $g^{I}_{y}$ denotes the symmetrised version of the expression under the sum in (I), by similar reasoning as above we can show that its envelope satisfies:
\begin{align*}
P^3(G^{I})^2 &\cong h^{-8} E[P(|Y_1-Y_2|<2h, |Y_1-Y_3|<2h, |Y_2-Y_3|<2h|X) K((X_{32}-X_{21})'\beta/h)]  = O(h^{-5}),
\end{align*}
and we also have: $P(\pi_{2,3}^P G^{I})^2 = O(h^{-2})$, which by Lemmas  \ref{lem:SZ}-\ref{lem:Gh} and our rate conditions implies:
\begin{gather}
|U_{n}^{(3)}g_{y}^{I} - P^3 g_{y}^{I} - 3 P_n \pi_{1,3} g_{y}^{I}| =  O_p(n^{-1}h^{-3/2} + n^{-3/2}h^{-5/2}) = o_p((nh)^{-1/2}). \label{eq:Idcmp}
\end{gather}
uniformly over $y$.

Finally, \eqref{eq:IIIfinal}, \eqref{eq:IIfinal} and \eqref{eq:Idcmp} give the result of the theorem and the expression for $\phi_i$ follows by direct calculation.

\section{Proof of Theorem \ref{thm:boot}}

For the global statistic, showing that $\sqrt{n}(S_{n}^{*}-\theta)$ converges weakly to a normal random variable involves standard arguments so we skip it. Then, using this result and Theorem \ref{thm:global},
we obtain the first result in Theorem \ref{thm:boot}. 

For the local test we have the decomposition:
\scriptsize
\begin{align}
U&_n^*(y) = \frac{1}{n(n-1)(n-2)}\bigg\{\sum_{i\neq j \neq k} \mathbbm{1}\{Y_{i}^*<Y_{j}^*<Y_{k}^* \} sgn(Y_{k}^*-2Y_{j}^*+Y_{i}^*)K_{h}(\mathbf{Y}^*-y)K_{h}\left((X_{kj} - X_{ji})'\hat{\beta}\right) &\text{(I)} \nonumber\\ 
& + \sum_{i\neq j \neq k} \mathbbm{1}\{Y_{i}^*<Y_{j}^*<Y_{k}^* \} sgn(Y_{k}^*-2Y_{j}^*+Y_{i}^*)K_{h}(\mathbf{Y}^*-y)K'_{h}\left((X_{kj} - X_{ji})'\hat{\beta}\right)(X_{kj} - X_{ji})'(\beta^*-\hat{\beta}) &\text{(II)} \nonumber\\ 
& + (\beta^*-\hat{\beta})'\sum_{i\neq j \neq k} \mathbbm{1}\{Y_{i}^*<Y_{j}^*<Y_{k}^* \} sgn(Y_{k}^*-2Y_{j}^*+Y_{i}^*)K_{h}(\mathbf{Y}^*-y)K''_{h}\left((X_{kj} - X_{ji})'\tilde{\beta}\right)(X_{kj} - X_{ji})(X_{kj} - X_{ji})'(\beta^* - \hat{\beta})\bigg\} &\text{(III)} \label{Bdecomp}
\end{align}
\normalsize
where $Y_i^* = X_i'\hat{\beta} + \eps^*_i$ and $\tilde{\beta}$ is between $\hat{\beta}$ and $\beta^*$.

We assume that $\eps^*_i$'s are drawn using the parametric bootstrap. The proof for wild bootstrap follows similar lines. The proof consists of three steps:
\begin{enumerate}
\item[Step 1.] Approximate $U_n^*(y)$ by $U_n^{\#}(y)$, where the latter statistic uses $Y_i^{\#} = X_i'\beta + \eps^{\#}_i$ and $\eps^{\#}_i$ is a bootstrap draw from $\tilde{\boldsymbol{\eps}} = \{\tilde{\eps}_i\}_{i=1}^n$ with $\tilde{\eps}_i = Y_i -X_i'\beta$.
\item[Step 2.] Show that a linear representation equivalent to the one in Theorem \ref{thm:local} holds for $U_n^{\#}(y)$ and approximate the test statistic $\sup_y  \sqrt{nh} U_n^{\#}(y)$, conditional on  $\{\tilde{\eps}_i\}_{i=1}^n$, by a Gaussian process.
\item[Step 3.] Approximate the sample statistic by a Gaussian process and show that, unconditionally, the difference between this process and the process in Step 2 is negligible. 
\end{enumerate}

\textbf{Step 1.} We will use Lemma \ref{lem:SZ} \eqref{lem:SZb} for
$U_n^{(3)} g^{I, diff}_{y, \hat{\beta}}, U_n^{(4)} g^{II, diff}_{y, \hat{\beta}}$ and $U_n^{(3)} g^{III, diff}_{y, \hat{\beta}}$ where:
\scriptsize
\begin{align*}
g^{I, diff}_{y, \hat{\beta}} =& \mathbbm{1}\{X_i'\hat{\beta} + \eps_i^*<X_j'\hat{\beta} + \eps_j^*<X_k'\hat{\beta} + \eps_k^* \} sgn(X_k'\hat{\beta} + \eps_k^*-2(X_j'\hat{\beta} + \eps_j^*)+X_i'\hat{\beta} + \eps_i^*)K_{h}(\mathbf{Y}^*-y)K_{h}\left((X_{kj} - X_{ji})'\hat{\beta}\right) \\
&- \mathbbm{1}\{X_i'\beta + \eps_i^{\#}<X_j'\beta + \eps_j^{\#}<X_k'\beta + \eps_k^{\#} \} sgn(X_k'\beta + \eps_k^{\#}-2(X_j'\beta + \eps_j^{\#})+X_i'\beta + \eps_i^{\#})K_{h}(\mathbf{Y}^{\#}-y)K_{h}\left((X_{kj} - X_{ji})'\beta\right)\\
g^{II, diff}_{y, \hat{\beta}} = & \mathbbm{1}\{X_i'\hat{\beta} + \eps_i^*<X_j'\hat{\beta} + \eps_j^*<X_k'\hat{\beta} + \eps_k^* \} sgn(X_k'\hat{\beta} + \eps_k^*-2(X_j'\hat{\beta} + \eps_j^*)+X_i'\hat{\beta} + \eps_i^*)K_{h}(\mathbf{Y}^*-y)K'_{h}\left((X_{kj} - X_{ji})'\hat{\beta}\right) \\
& \qquad \qquad \qquad \qquad \qquad \qquad \qquad \qquad \qquad \qquad \qquad \qquad \qquad \qquad \qquad \qquad \qquad \qquad \qquad  \times (X_{kj} - X_{ji})' \Omega(X_l, X_l'\hat{\beta}+  \eps_l^*) \\
&- \mathbbm{1}\{X_i'\beta + \eps_i^{\#}<X_j'\beta + \eps_j^{\#}<X_k'\beta + \eps_k^{\#} \} sgn(X_k'\beta + \eps_k^{\#}-2(X_j'\beta + \eps_j^{\#})+X_i'\beta + \eps_i^{\#})K_{h}(\mathbf{Y}^{\#}-y)K'_{h}\left((X_{kj} - X_{ji})'\beta\right) \\
& \qquad \qquad \qquad \qquad \qquad \qquad \qquad \qquad \qquad \qquad \qquad \qquad \qquad \qquad \qquad \qquad \qquad \qquad \qquad  \times (X_{kj} - X_{ji})' \Omega(X_l, X_l'\beta+  \eps_l^{\#})\\
g^{III, diff}_{y, \hat{\beta}} = & \mathbbm{1}\{X_i'\hat{\beta} + \eps_i^*<X_j'\hat{\beta} + \eps_j^*<X_k'\hat{\beta} + \eps_k^* \} sgn(X_k'\hat{\beta} + \eps_k^*-2(X_j'\hat{\beta} + \eps_j^*)+X_i'\hat{\beta} + \eps_i^*)K_{h}(\mathbf{Y}^*-y)K''_{h}\left((X_{kj} - X_{ji})'\hat{\beta}\right) \\
& \qquad \qquad \qquad \qquad \qquad \qquad \qquad \qquad \qquad \qquad \qquad \qquad \qquad \qquad \qquad \qquad \qquad \qquad \qquad  \times (X_{kj} - X_{ji})(X_{kj} - X_{ji})' \\
&- \mathbbm{1}\{X_i'\beta + \eps_i^{\#}<X_j'\beta + \eps_j^{\#}<X_k'\beta + \eps_k^{\#} \} sgn(X_k'\beta + \eps_k^{\#}-2(X_j'\beta + \eps_j^{\#})+X_i'\beta + \eps_i^{\#})K_{h}(\mathbf{Y}^{\#}-y)K''_{h}\left((X_{kj} - X_{ji})'\beta\right) \\
& \qquad \qquad \qquad \qquad \qquad \qquad \qquad \qquad \qquad \qquad \qquad \qquad \qquad \qquad \qquad \qquad \qquad \qquad \qquad  \times(X_{kj} - X_{ji})(X_{kj} - X_{ji})'
\end{align*}
\normalsize
Note that the functions above are indexed both by $y$ and $\beta$. In order to apply the lemma we will show that as $\|\hat{\beta} - \beta\|/h \to 0$ the variance of these functions converges to zero, uniformly over $y\in \mathcal{Y}$. 

Let $m_i \sim Multinomial(1; \frac{1}{n},\ldots,\frac{1}{n})$.  We can write $\eps_i^* = \hat{\eps}'m_i = (\tilde{\eps} -\mathbf{X}(\hat{\beta}-\beta))'m_i $ and $\eps^{\#}_i = \tilde{\eps}'m_i$. 
Firstly, we have:
\scriptsize
\begin{align*}
E&[\mathbbm{1}\{X_i'\hat{\beta} + \eps_i^*<X_j'\hat{\beta} + \eps_j^*<X_k'\hat{\beta} + \eps_k^* \} sgn(X_k'\hat{\beta} + \eps_k^*-2(X_j'\hat{\beta} + \eps_j^*)+X_i'\hat{\beta} + \eps_i^*)K_{h}(\mathbf{Y}^*-y)|\hat{\beta},\mathcal{X},m_i,m_j, m_k]
= \\
&\int E[\mathbbm{1}\{X_i'\hat{\beta} + (\tilde{\boldsymbol{\eps}} -\mathbf{X}(\hat{\beta}-\beta))'m_i<X_j'\hat{\beta} + (\tilde{\boldsymbol{\eps}} -\mathbf{X}(\hat{\beta}-\beta))'m_j<X_k'\hat{\beta} + (\tilde{\boldsymbol{\eps}} -\mathbf{X}(\hat{\beta}-\beta))'m_k \} sgn(X_k'\hat{\beta} + (\tilde{\boldsymbol{\eps}} -\mathbf{X}(\hat{\beta}-\beta))'m_k\\
&-2(X_j'\hat{\beta} + (\tilde{\boldsymbol{\eps}} -\mathbf{X}(\hat{\beta}-\beta))'m_j)+X_i'\hat{\beta} + (\tilde{\boldsymbol{\eps}} -\mathbf{X}(\hat{\beta}-\beta))'m_i)K_{h}\left(\begin{pmatrix} X_i'\hat{\beta} + (\tilde{\boldsymbol{\eps}} -\mathbf{X}(\hat{\beta}-\beta))'m_i \\ X_j'\hat{\beta} + (\tilde{\boldsymbol{\eps}} -\mathbf{X}(\hat{\beta}-\beta))'m_j \\ X_k\hat{\beta} + (\tilde{\boldsymbol{\eps}} -\mathbf{X}(\hat{\beta}-\beta))'m_k \end{pmatrix}-y\right) f_{\tilde{\eps}'m}\left(\begin{pmatrix} \tilde{\boldsymbol{\eps}}'m_i \\ \tilde{\boldsymbol{\eps}}'m_j \\ \tilde{\boldsymbol{\eps}}'m_k  \end{pmatrix}{\Bigg |} \mathcal{X}\right) d\tilde{\eps} \\
&= \int \mathbbm{1}\{s_1<s_2<s_3\}sgn\{s_3-2s_2+s_1\}K(s_1)K(s_2)K(s_3)  f_{\tilde{\eps}'m}\left(\begin{pmatrix} s_1h +y - X_i'\hat{\beta} + (\mathbf{X}(\hat{\beta}-\beta))'m_i\ \\ s_2h +y - X_j'\hat{\beta} + (\mathbf{X}(\hat{\beta}-\beta))'m_j \\ s_3h +y - X_k'\hat{\beta} + (\mathbf{X}(\hat{\beta}-\beta))'m_k \end{pmatrix}{\Bigg |} \mathcal{X}\right) ds 
\end{align*}
\normalsize
and $h K_{h}\left((X_{kj} - X_{ji})'\hat{\beta}\right) = K\left( (X_{kj} - X_{ji})'\beta /h \right) + O(\|\hat{\beta}-\beta\|/h)$. Similarly:
\scriptsize
\begin{align*}
E&[\mathbbm{1}\{X_i'\beta + \eps_i^{\#}<X_j'\beta + \eps_j^{\#}<X_k'\beta + \eps_k^{\#} \} sgn(X_k'\beta + \eps_k^{\#}-2(X_j'\beta + \eps_j^{\#})+X_i'\beta + \eps_i^{\#})K_{h}(\mathbf{Y}^{\#}-y)|\mathcal{X},m_i,m_j, m_k]
= \\
& = \int \mathbbm{1}\{s_1<s_2<s_3\}sgn\{s_3-2s_2+s_1\}K(s_1)K(s_2 )K(s_3)  f_{\tilde{\eps}'m}\left(\begin{pmatrix} s_1h +y - X_i'\beta \\ s_2h +y - X_j'\beta \\ s_3h +y - X_k'\hat{\beta} \end{pmatrix}{\Bigg |} \mathcal{X}\right) ds 
\end{align*}
\normalsize
Thus, using 
$f(y+sh) = f(y) + O(sh)$, Lemma \ref{lem:int} and:
\footnotesize
\begin{align*}
 f_{\tilde{\eps}'m}&\left(\begin{pmatrix} s_1h +y - X_i'\hat{\beta} + (\mathbf{X}(\hat{\beta}-\beta))'m_i\ \\ s_2h +y - X_j'\hat{\beta} + (\mathbf{X}(\hat{\beta}-\beta))'m_j \\ s_3h +y - X_k'\hat{\beta} + (\mathbf{X}(\hat{\beta}-\beta))'m_k \end{pmatrix}  {\Bigg |} \mathcal{X}\right) = \\
 &=  f_{\tilde{\eps}'m}\left(\begin{pmatrix} s_1h +y - X_i'\beta \\ s_2h +y - X_j'\beta \\ s_3h +y - X_k'\beta \end{pmatrix}  {\Bigg |} \mathcal{X}\right) + \nabla  f_{\tilde{\eps}'m}\left(\begin{pmatrix} s_1h +y - X_i'\beta \\ s_2h +y - X_j'\beta \\ s_3h +y - X_k'\beta \end{pmatrix}  {\Bigg |} \mathcal{X}\right)'\begin{pmatrix} m_i'\mathbf{X} -X_i' \\ m_j'\mathbf{X} - X_j' \\ m_k'\mathbf{X} - X_k' \end{pmatrix} (\hat{\beta}-\beta) + O(\|(\hat{\beta}-\beta)\|^2)
\end{align*}
\normalsize
we obtain:
\begin{gather*}
\sup_y E\left[g^{I, diff}_{y, \hat{\beta}}\right]  = O(\|\hat{\beta}-\beta\| + h \|\hat{\beta}-\beta\| + \|(\hat{\beta}-\beta)\|^2) = o((nh)^{-1/2}) 
 \end{gather*}
 This allows us to define a P-canonical function $\tilde{g}^{I,diff}_{y, \hat{\beta}} =  g^{I, diff}_{y, \hat{\beta}} -  E\left[g^{I, diff}_{y, \hat{\beta}}\right]$ and write $U_n^{(3)} g^{I, diff}_{y, \hat{\beta}} = U_n^{(3)} \tilde{g}^{I, diff}_{y, \hat{\beta}} + o_p((nh)^{-1/2})$. 
 
 Furthermore, by similar calculation:
 \scriptsize
 \begin{align*}
 E&[\mathbbm{1}\{X_i'\hat{\beta} + \eps_i^*<X_j'\hat{\beta} + \eps_j^*<X_k'\hat{\beta} + \eps_k^* \} sgn^2(X_k'\hat{\beta} + \eps_k^*-2(X_j'\hat{\beta} + \eps_j^*)+X_i'\hat{\beta} + \eps_i^*)K^2_{h}(\mathbf{Y}^*-y)|\hat{\beta},\mathcal{X},m_i,m_j, m_k]
= \\
&= h^{-3} \int \mathbbm{1}\{s_1<s_2<s_3\}sgn^2\{s_3-2s_2+s_1\}K^2(s_1)K^2(s_2)K^2(s_3)  f_{\tilde{\eps}'m}\left(\begin{pmatrix} s_1h +y - X_i'\hat{\beta} - \mathbf{X}(\hat{\beta}-\beta)'m_i\ \\ s_2h +y - X_j'\hat{\beta} - \mathbf{X}(\hat{\beta}-\beta)'m_j \\ s_3h +y - X_k'\hat{\beta} - \mathbf{X}(\hat{\beta}-\beta)'m_k \end{pmatrix}{\Bigg |} \mathcal{X}\right) ds 
 \end{align*}
\normalsize
and 
\scriptsize
 \begin{align*}
 E&[\mathbbm{1}\{X_i'\hat{\beta} + \eps_i^*<X_j'\hat{\beta} + \eps_j^*<X_k'\hat{\beta} + \eps_k^* \} sgn(X_k'\hat{\beta} + \eps_k^*-2(X_j'\hat{\beta} + \eps_j^*)+X_i'\hat{\beta} + \eps_i^*)K_{h}(\mathbf{Y}^*-y) \\
 & \quad \times \mathbbm{1}\{X_i'\beta + \eps_i^{\#}<X_j'\beta + \eps_j^{\#}<X_k'\beta + \eps_k^{\#} \} sgn(X_k'\beta + \eps_k^{\#}-2(X_j'\beta + \eps_j^{\#})+X_i'\beta + \eps_i^{\#})K_{h}(\mathbf{Y}^{\#}-y)|\hat{\beta}, \mathcal{X},m_i,m_j, m_k]
= \\
&= h^{-3} \int \mathbbm{1}\{s_1<s_2<s_3\} \mathbbm{1}\{s_1 + X_i'(\hat{\beta}-\beta)/h +  (\mathbf{X}(\hat{\beta}-\beta))'m_i/h < s_2 + X_j'(\hat{\beta}-\beta)/h + (\mathbf{X}(\hat{\beta}-\beta))'m_j/h \\
&\qquad \qquad \qquad \qquad \qquad \qquad \qquad \qquad \qquad \qquad \qquad \qquad \qquad \qquad \qquad \qquad \qquad \qquad  < s_3 + X_k'(\hat{\beta}-\beta)/h + (\mathbf{X}(\hat{\beta}-\beta))'m_k/h\} \\
& \quad \times sgn\{s_3-2s_2+s_1\}  sgn\{s_3  + X_k'(\hat{\beta}-\beta)/h + (\mathbf{X}(\hat{\beta}-\beta))'m_k/h -2s_2  -2X_j'(\hat{\beta}-\beta)/h-2(\mathbf{X}(\hat{\beta}-\beta))'m_j/h\\
& \qquad \qquad \qquad \qquad \qquad \qquad \qquad \qquad \qquad \qquad \qquad \qquad \qquad \qquad \qquad \qquad \qquad \qquad + s_1 +X_i'(\hat{\beta}-\beta)/h +  (\mathbf{X}(\hat{\beta}-\beta))'m_i/h\} \\
& \quad \times K^2(s_1)K^2(s_2)K^2(s_3)  f_{\tilde{\eps}'m}\left(\begin{pmatrix} s_1h +y - X_i'\beta \ \\ s_2h +y - X_j'\beta  \\ s_3h +y - X_k'\beta \end{pmatrix}{\Bigg |} \mathcal{X}\right) ds + O(h^{-4} \|\hat{\beta} -\beta\|) \
 \end{align*}
\normalsize
which again implies that:
\begin{gather*}
 \sup_y E\left[h^4 \left(\tilde{g}^{I, diff}_{y, \hat{\beta}}\right)^2\right] \to 0 
\end{gather*}
 as $\|\hat{\beta} - \beta\|/h \to 0$. Finally, by Lemma \ref{lem:Eucl} the function class $\{\tilde{g}^{I, diff}_{y, \beta}: y \in \mathcal{Y}, \beta \in \Theta_{\beta}\}$ is Euclidean and we can apply Lemma \ref{lem:SZ}\eqref{lem:SZb} to show that:
\begin{gather*}
\sup_y \sup_{\|\hat{\beta} - \beta\|/h<\delta_n} | U_n^{(3)} g^{I, diff}_{y, \hat{\beta}} | = o_p(n^{-3/2} h^{-2})  +  o_p((nh)^{-1/2})= o_p((nh)^{-1/2})
\end{gather*}
for any $\delta_n \to 0$. 

Further, we have $E[g^{II, diff}_{y, \hat{\beta}}]=0$, $h^2 K'_{h}\left((X_{kj} - X_{ji})'\hat{\beta}\right) = K'\left( (X_{kj} - X_{ji})'\beta /h \right) + O(\|\hat{\beta}-\beta\|/h)$ and $E[((X_{kj}-X_{ji})'\Omega(X_l,X_l'\beta+\eps_l^{\#}))^2]<\infty$ using our assumption on the moments of $X$ and $\Omega$.  Thus, using continuity of $\Omega$ (see Assumption \ref{BA1d}), dominated convergence theorem and similar reasoning as above, we obtain $\sup_y E\left[h^7 \left(g^{II, diff}_{y, \hat{\beta}}\right)^2\right] \to 0$ as $\|\hat{\beta} - \beta\|/h \to 0$, which by Lemma \ref{lem:SZ}\eqref{lem:SZb} implies: 
\begin{align*}
\sup_y \sup_{\|\hat{\beta} - \beta\|/h<\delta_n} | U_n^{(4)} \tilde{g}^{II, diff}_{y, \hat{\beta}} | = o_p(n^{-2} h^{-7/2}) = o_p((nh)^{-1/2})
\end{align*}   

By similar reasoning as for $g^{I,diff}_{y,\hat{\beta}}$, we have $\sup_y E[g^{III, diff}_{y, \hat{\beta}}] = o((nh)^{-1/2})$ and $E\left[h^9 \left(\tilde{g}^{III, diff}_{y, \hat{\beta}}\right)^2\right] \to 0$ uniformly over $y$, where $\tilde{g}^{III, diff}_{y, \hat{\beta}}$ is a recentred version of $g^{III, diff}_{y, \hat{\beta}}$, which by Lemma  \ref{lem:SZ}\eqref{lem:SZb} implies:
\begin{align*}
\sup_y \sup_{\|\hat{\beta} - \beta\|/h<\delta_n} | U_n^{(3)} g^{III, diff}_{y, \hat{\beta}}| = o_p(n^{-3/2} h^{-9/2}) + o_p((nh)^{-1/2}).
\end{align*} 
Together with $\beta^*-\hat{\beta} = O_p^*(n^{-1/2}) =  O_p(n^{-1/2})$ this implies:
\begin{gather*}
\sup_y \sup_{\|\hat{\beta} - \beta\|/h<\delta_n} |(\beta^*-\hat{\beta})'U_n^{(3)} g^{III, diff}_{y, \hat{\beta}}(\beta^*-\hat{\beta})| = o_p(n^{-5/2} h^{-9/2}) = o_p((nh)^{-1/2}).
\end{gather*}

Finally, using Assumption \ref{BA1d} and putting all the results above together we obtain:
\begin{gather}
U_n^*(y) = U_n^{\#} g_y^{I} + U_n^{\#} g_y^{II} + (\beta^*-\hat{\beta})' U_n^{\#} g_y^{III} (\beta^*-\hat{\beta}) + o_p((nh)^{-1/2}). \label{eq:hashboot}
\end{gather}

\textbf{Step 2.} We will apply the second part of Lemma \ref{lem:SZ}\eqref{lem:SZa} and Lemma \ref{lem:Gh} in the same fashion as in the proof of Theorem \ref{thm:local}, the only difference being that now the error terms are drawn from the distribution of $\tilde{\eps}$ where $\tilde{\eps} = Y - X'\beta$, but still independently of $X$. 
Also note that:
\scriptsize
\begin{align*}
E& \mathbbm{1}\{X_i'\beta + \eps_i^{\#}<X_j'\beta + \eps_j^{\#}<X_k'\beta + \eps_k^{\#} \} sgn(X_k'\beta + \eps_k^{\#}-2(X_j'\beta + \eps_j^{\#})+X_i'\beta + \eps_i^{\#})K_{h}(\mathbf{Y}^{\#}-y)|X_i'\beta, X_j'\beta, X_k'\beta]= \\
&= E \left[ \mathbbm{1}\{X_i'\beta + \tilde{\boldsymbol{\eps}}'m_i<X_j'\beta + \tilde{\boldsymbol{\eps}}' m_j<X_k'\beta + \tilde{\boldsymbol{\eps}}' m_k \} sgn(X_k'\beta + \tilde{\boldsymbol{\eps}}'m_k-2(X_j'\beta + \tilde{\boldsymbol{\eps}}' m_j)+X_i'\beta + \tilde{\boldsymbol{\eps}}' m_i)  \vphantom{\begin{pmatrix} X_i'\beta + \tilde{\boldsymbol{\eps}}'m_i \\ X_j'\beta + \tilde{\boldsymbol{\eps}}'m_j \\ X_k'\beta + \tilde{\boldsymbol{\eps}}'m_k  \end{pmatrix}} \right. \\
& \left. \qquad \qquad \qquad \qquad \qquad \qquad \qquad \qquad \qquad \qquad \qquad \qquad \qquad \qquad \qquad \qquad \times K_{h}\left(\begin{pmatrix} X_i'\beta + \tilde{\boldsymbol{\eps}}'m_i \\ X_j'\beta + \tilde{\boldsymbol{\eps}}'m_j \\ X_k'\beta + \tilde{\boldsymbol{\eps}}'m_k  \end{pmatrix} -y\right)\Bigg|X_i'\beta, X_j'\beta, X_k'\beta \right]=\\
& = E\left[\frac{1}{n^3} \sum_{l,p,q} \mathbbm{1}\{X_i'\beta + \tilde{\eps}_l<X_j'\beta + \tilde{\eps}_p<X_k'\beta + \tilde{\eps}_q \} sgn(X_k'\beta + \tilde{\eps}_q - 2(X_j'\beta + \tilde{\eps}_p)+X_i'\beta + \tilde{\eps}_l)  \vphantom{\begin{pmatrix} X_i'\beta + \tilde{\boldsymbol{\eps}}'m_i \\ X_j'\beta + \tilde{\boldsymbol{\eps}}'m_j \\ X_k'\beta + \tilde{\boldsymbol{\eps}}'m_k  \end{pmatrix}} \right. \\
& \left. \qquad \qquad \qquad \qquad \qquad \qquad \qquad \qquad \qquad \qquad \qquad \qquad \qquad \qquad \qquad \qquad \times K_{h}\left(\begin{pmatrix} X_i'\beta + \tilde{\eps}_l \\ X_j'\beta + \tilde{\eps}_p \\ X_k'\beta + \tilde{\eps}_q  \end{pmatrix} -y\right)\Bigg|X_i'\beta, X_j'\beta, X_k'\beta \right]= \\
& = E\left[ \mathbbm{1}\{X_i'\beta + \tilde{\eps}_l<X_j'\beta + \tilde{\eps}_p<X_k'\beta + \tilde{\eps}_q \} sgn(X_k'\beta + \tilde{\eps}_q - 2(X_j'\beta + \tilde{\eps}_p)+X_i'\beta + \tilde{\eps}_l)  \vphantom{\begin{pmatrix} X_i'\beta + \tilde{\boldsymbol{\eps}}'m_i \\ X_j'\beta + \tilde{\boldsymbol{\eps}}'m_j \\ X_k'\beta + \tilde{\boldsymbol{\eps}}'m_k  \end{pmatrix}} \right. \\
& \left.  \qquad \qquad \qquad \qquad \qquad \qquad \qquad \qquad \qquad \qquad \qquad \qquad \qquad \qquad \qquad \times K_{h}\left(\begin{pmatrix} X_i'\beta + \tilde{\eps}_l \\ X_j'\beta + \tilde{\eps}_p \\ X_k'\beta + \tilde{\eps}_q  \end{pmatrix} -y\right)\Bigg|X_i'\beta, X_j'\beta, X_k'\beta \right] + O(n^{-1})
\end{align*}
\normalsize
which gives $E[U_n^{\#} g_y^{I}] = \tilde{P}^3  g_y^{I} + O(n^{-1})$, where $\tilde{P}$ is the probability measure under which $Y_i = X_i\beta + \tilde{\eps}_i$ and $\tilde{\eps}_i$ is drawn independently of $X_i$ from $f_{\tilde{\eps}}$. 

This gives, uniformly over $y$:
\begin{align*}
(\beta^*-\hat{\beta})' U_n^{\#} g_y^{III} (\beta^*-\hat{\beta}) &= O_{p}(n^{-1}h + n^{-3/2}h^{-7/2} + n^{-2}h^{-9/2}) \\
U_n^{\#} g_y^{II} & = O_p(n^{-1/2}h + n^{-1}h^{-1} + n^{-3/2} h^{-3/2} +n^{-2}h^{-7/2}) \\
U_n^{\#} g_y^{I} &= \tilde{P}^3 g_{y}^{I} + 3 P_n^{\#} \pi_{1,3}^{\tilde{P}} g_{y}^{I} +  O_p(n^{-1}h^{-3/2} + n^{-3/2}h^{-5/2})
\end{align*}
and therefore:
\begin{gather}
U_n^*(y) = \tilde{P}^3 g_{y}^{I} + 3 P_n^{\#} \pi_{1,3}^{\tilde{P}}  g_{y}^{I} + o_p((nh)^{-1/2}). \label{eq:bootlin}
\end{gather}

Let $E_{|\boldsymbol{\tilde{\eps}}}$ denote the expectation conditional on $\boldsymbol{\tilde{\eps}}$. We will apply Lemma \ref{lem:Ch} conditionally on $\boldsymbol{\tilde{\eps}}$ with $\tilde{B}(y) = \sqrt{nh} \tilde{P}^3 g_{y}^{I}$ and $ \mathbbm{G}_P \phi_n^{\boldsymbol{\tilde{\eps}}}(y)$ where:\footnote{Note that the results in \cite{chernozhukov_et_al16} do not apply directly to our parametric bootstrap.}   
\footnotesize
\begin{align*}
\phi_n^{\boldsymbol{\tilde{\eps}}}(y) =& 3\sqrt{h} E_{|\boldsymbol{\tilde{\eps}}}[\mathbbm{1}\{X_i'\beta + \boldsymbol{\tilde{\eps}}'m_i <X_j'\beta + \boldsymbol{\tilde{\eps}}'m_j <X_k'\beta + \boldsymbol{\tilde{\eps}}' m_k \} sgn(X_k'\beta + \boldsymbol{\tilde{\eps}}' m_k-2(X_j'\beta + \boldsymbol{\tilde{\eps}}' m_j)+X_i'\beta + \boldsymbol{\tilde{\eps}}' m_i)\\
&  \qquad \qquad \qquad \qquad \qquad \qquad \qquad \qquad \qquad \qquad \times K_{h}(\mathbf{Y}^{\#}-y) K_h((X_{kj}-X_{ji})'\beta)|X_i, m_i] + \textit{symmetric terms}.
\end{align*}
\normalsize
Let us first verify conditions of Lemma \ref{lem:Ch}. Condition \eqref{lem:ChA} is satisfied because $\tilde{P}^3 g_{y}^{I}$ is a continuous function of $y \in \mathcal{Y}$  where $\mathcal{Y}$ is a compact set. Condition \eqref{lem:ChB} is satisfied because  $\{\phi_n^{\boldsymbol{\tilde{\eps}}}(y): y\in \mathcal{Y}\}$ as a class of functions of $(X_i, m_i)$ is Euclidean by the same arguments as those leading to Lemma \ref{lem:Eucl} (note that this class depends on $n$ both through $h$ and $\boldsymbol{\tilde{\eps}}$).   

Next, in order to show that condition \eqref{lem:ChC}  of Lemma \ref{lem:Ch} is satisfied, it is enough to verify it for $\sqrt{h}\phi_n^{\boldsymbol{\tilde{\eps}}}(y)$. Define:  
\footnotesize
\begin{align*}
G_{h,y}^{\boldsymbol{\tilde{\eps}}}(X_{i}'\beta, \boldsymbol{\tilde{\eps}}'m_i) \equiv & 3 E_{|\boldsymbol{\tilde{\eps}}}[\mathbbm{1}\{X_{i}'\beta +  \boldsymbol{\tilde{\eps}}'m_i< X_{j}'\beta +  \boldsymbol{\tilde{\eps}}'m_j < X_{k}'\beta +  \boldsymbol{\tilde{\eps}}'m_k \} sgn(X_{k}'\beta +  \boldsymbol{\tilde{\eps}}'m_k - 2(X_{j}'\beta +  \boldsymbol{\tilde{\eps}}'m_j)+X_{i}'\beta +  \boldsymbol{\tilde{\eps}}'m_i) \\ 
& \times K_{h}(X_{j}'\beta +  \boldsymbol{\tilde{\eps}}'m_j-y)K_{h}(X_{k}'\beta +  \boldsymbol{\tilde{\eps}}'m_k-y)K_{h}\left((X_{kj} - X_{ji})'\beta\right)| X_{i}'\beta, m_i] + \textit{symmetric terms}
\end{align*}
\normalsize
and note that $\sqrt{h} \phi_n^{\boldsymbol{\tilde{\eps}}}(y) = G_{h,y}^{\boldsymbol{\tilde{\eps}}}(X_{i}'\beta, \boldsymbol{\tilde{\eps}}'m_i)K((X_{i}'\beta +  \boldsymbol{\tilde{\eps}}'m_i - y)/h)$. Furthermore, using smoothness and boundedness of the distribution of $X_j'\beta, $ (see Assumption \ref{A1}\eqref{A1c}), by change of variables and dominated convergence theorem:
\scriptsize
\begin{align*}
E_{|\boldsymbol{\tilde{\eps}}}&[\mathbbm{1}\{X_{i}'\beta +  \boldsymbol{\tilde{\eps}}'m_i< X_{j}'\beta +  \boldsymbol{\tilde{\eps}}'m_j < X_{k}'\beta +  \boldsymbol{\tilde{\eps}}'m_k \} sgn(X_{k}'\beta +  \boldsymbol{\tilde{\eps}}'m_k - 2(X_{j}'\beta +  \boldsymbol{\tilde{\eps}}'m_j)+X_{i}'\beta +  \boldsymbol{\tilde{\eps}}'m_i)  K_{h}(X_{j}'\beta +  \boldsymbol{\tilde{\eps}}'m_j-y) \\
&  \qquad \qquad \qquad \qquad \qquad \qquad \qquad \qquad \qquad \qquad \qquad \qquad \qquad \qquad \qquad  \times K_{h}(X_{k}'\beta +  \boldsymbol{\tilde{\eps}}'m_k-y)|(X_{kj}-X_{ji})'\beta, X_i'\beta, m_i]  \\
& =\frac{1}{n^2} \sum_{p,q} \int_{-1}^1 \mathbbm{1}\{X_i'\beta + \boldsymbol{\tilde{\eps}}'m_i < s_2 h +y  < s_3 h + y  \} sgn(s_3 h + y  -2 (s_2 h + y ) + X_i'\beta + \boldsymbol{\tilde{\eps}}'m_i\}   K(s_2)K(s_3)  \\
&  \qquad \qquad \qquad \qquad \qquad \qquad \qquad \qquad \qquad  \qquad \times f_{xb,xb}(s_2 h + y -\tilde{\eps}_p, s_3 h + y - \tilde{\eps}_q  |(X_{kj}-X_{ji})'\beta, X_i'\beta, \boldsymbol{\tilde{\eps}}) ds = O(1)
\end{align*}
\normalsize
where $f_{xb,xb}(\cdot|(X_{kj}-X_{ji})'\beta, X_i'\beta, \boldsymbol{\tilde{\eps}})$ denotes conditional distribution of $(X_j'\beta, X_k'\beta)$. This implies $G_{h,y}^{\boldsymbol{\tilde{\eps}}}(X_{i}'\beta, \boldsymbol{\tilde{\eps}}'m_i) = O(1)$. 
Now we have:
\footnotesize
\begin{align*}
E_{|\boldsymbol{\tilde{\eps}}}\left[(G_{h,y}^{\boldsymbol{\tilde{\eps}}}(X_{i}'\beta, \boldsymbol{\tilde{\eps}}'m_i))^{2} K^2\left(\frac{X_{i}'\beta +  \boldsymbol{\tilde{\eps}}'m_i -y}{h}\right)\right] 
= h  \frac{1}{n} \sum_{l} \int (G_{h,y}^{\boldsymbol{\tilde{\eps}}}(uh+y - \tilde{\eps}_l,\tilde{\eps}_{l}))^{2}K^{2}\left(u\right) f_{xb}(uh+y-\tilde{\eps}_l)du =O(h)
\end{align*} 
\normalsize
uniformly over $y$ and by similar reasoning $\sup_y E_{|\boldsymbol{\tilde{\eps}}} [h^{3/2} \phi_n^{\boldsymbol{\tilde{\eps}}}(y)^3] = O(h)$ and  $\sup_y E_{|\boldsymbol{\tilde{\eps}}} [h^2 \phi_n^{\boldsymbol{\tilde{\eps}}}(y)^4] = O(h)$. The envelope of $\phi_n^{\boldsymbol{\tilde{\eps}}}(y)$ is given by:
\footnotesize
\begin{align}
 G^{\boldsymbol{\tilde{\eps}}} =& h^{-7/2} \frac{1}{n^2} \sum_{p,q} E_{|\boldsymbol{\tilde{\eps}}}[\mathbbm{1}\{|X_{21}'\beta + \tilde{\eps}_p - \boldsymbol{\tilde{\eps}}'m_1|<2h\} \mathbbm{1}\{|X_{31}'\beta + \tilde{\eps}_q - \boldsymbol{\tilde{\eps}}'m_1|<2h\}  \mathbbm{1}\{|X_{32}'\beta + \tilde{\eps}_q - \tilde{\eps}_p|<2h\} \nonumber \\ 
 & \times K((X_{32}-X_{21})'\beta/h)|X_1'\beta, m_1 ] = O(h^{-1/2}) \label{eq:condenvlp}
\end{align}
\normalsize
where the last equality follows from $P(|X_{21}'\beta + \tilde{\eps}_p - \boldsymbol{\tilde{\eps}}'m_1|<2h, |X_{31}'\beta + \tilde{\eps}_q - \boldsymbol{\tilde{\eps}}'m_1|<2h, |X_{32}'\beta + \tilde{\eps}_q - \tilde{\eps}_p|<2h| X, \boldsymbol{\tilde{\eps}}) = O(h^2)$ and the standard change of variables argument. This further implies that $\| \sqrt{h}   G^{\boldsymbol{\tilde{\eps}}}\|_{P,q} = O(1)$ for $q=2,3,4$.

Thus, we can apply Lemma \ref{lem:Ch} with $q=4, K_n=\log n,\gamma = (\log n)^{-1}, \sigma^2=O(h)$ and $b=O(1)$  to obtain that there exists a random variable $Z^{\boldsymbol{\tilde{\eps}}}$ such that $\sqrt{h}|\sup_{y} \{\tilde{B}(y) + \sqrt{nh}P_{n}^{\#}\phi_n^{\boldsymbol{\tilde{\eps}}}(y)\} -Z^{\boldsymbol{\tilde{\eps}}}| = O_{p}(n^{-1/6}h^{1/3} \log n + n^{-1/4}\log^{5/4} n)$ and $Z^{\boldsymbol{\tilde{\eps}}}$ follows the same distribution as $\sup_{y} \{\tilde{B}(y) + \mathbbm{G}_P \phi_n^{\boldsymbol{\tilde{\eps}}}(y)\}$, which implies:
\begin{align}
|\sup_{y} \{\tilde{B}(y) + \sqrt{nh}P_{n}^{\#}\phi_n^{\boldsymbol{\tilde{\eps}}}(y)\} -Z^{\boldsymbol{\tilde{\eps}}}| &= O_{p}(n^{-1/6}h^{-1/6} \log n + n^{-1/4}h^{-1/2}\log^{5/4} n) \nonumber \\
&= o_{p}(1) \label{eq:final1}
\end{align}
where the last equality follows from $nh^2/(\log n)^5 \to \infty$, which is implied by the rate condition in Assumption \ref{A1}\eqref{A1b}. 

\textbf{Step 3.} For the sample statistic, we have from Theorem \ref{thm:local}:
\begin{gather*}
U_{n}(y)  = P^3 g_{y}^{I} + 3 P_n \pi_{1,3}^P g_{y}^{I} + o_p((nh)^{-1/2}).
\end{gather*} 
uniformly over $y$. Denote $\phi_n(y) = 3 \sqrt{h}  \pi_{1,3}^P g_{y}^{I}$. Following similar, and in fact simpler, reasoning as above in Step 2 (e.g. now $\eps_i$'s are not fixed so we can integrate them out) under the null hypothesis Lemma \ref{lem:Ch} implies:
\begin{gather}
|\sup_{y} \{B(y) + \sqrt{n}P_{n}\phi_n(y)\} -Z| = o_p(1) \label{eq:final2}
\end{gather}
where $Z$ follows the same distribution as $\sup_{y} \{B(y) + \mathbbm{G}_P \phi_n(y)\}$.

Now $\mathbbm{G}_P \phi_n(y)$ and $\mathbbm{G}_P \phi_n^{\boldsymbol{\eps}}(y)$ are processes defined on the joint probability space $(\mathcal{W}^{\infty} \times \{0,1\}^{\infty}, \mathcal{A} \times \mathcal{C}, P \times P_M)$, where $P_M$ is the multinomial distribution. 
We have:
\small
\begin{align}
|\sup_y \sqrt{nh} U_n^*(y) -  \sup_y \{B(y) + \mathbbm{G}_P \phi_n(y) \} | \leq & |\sup_y \sqrt{nh} U_n^*(y) -  \sup_y \{B(y) +  \mathbbm{G}_P \phi_n^{\boldsymbol{\eps}}(y) \} | \nonumber \\
& \qquad \qquad \qquad \qquad  +  \sup_y | \mathbbm{G}_P \phi_n^{\boldsymbol{\eps}}(y) -  \mathbbm{G}_P \phi_n(y)| \label{eq:final3}
\end{align}
\normalsize
Define $\gamma_n(y) = \phi_n^{\boldsymbol{\eps}}(y) - \phi_n(y)$ and the corresponding class of functions by $\Gamma$. Corollary 2.2.8. in \cite{van_der_vaart_wellner96} implies that, if $ \mathbbm{G}_P \gamma_n(y)$ is a separable sub-Gaussian process:
\begin{gather*}
E  \left[\sup_y | \mathbbm{G}_P \phi_n^{\boldsymbol{\eps}}(y) -  \mathbbm{G}_P \phi_n(y)| \right] \lesssim \int_0^{diam/2} \sqrt{\log N(\varepsilon, \Gamma, L_2(P \times P_M)) } d \varepsilon
\end{gather*}
where $diam$ is the $L_2(P \times P_M)$ diameter of $\Gamma$. Firstly, note that $\mathbbm{G}_P \gamma_n(y)$ is a Gaussian process with $L_2(P \times P_M)$-continuous sample paths (see Lemma \ref{lem:Ch}) indexed by a compact set $\mathcal{Y}$ so it is a separable sub-Gaussian process for the $L_2(P \times P_M)$ semi-metric. 

We have:
\small
\begin{gather*}
E[ \gamma_n^2(y)]  = 9 h E\left[\frac{1}{n} \sum_j (\pi_{1,3}^P g_y^{I}(X_i'\beta, \eps_j) - \pi_{1,3}^P g_y^{I}(X_i'\beta, \eps_i))^2\right] = 9 h \frac{n-1}{n} E\left[(\pi_{1,3}^P g_y^{I}(X_i'\beta, \eps_j) - \pi_{1,3}^P g_y^{I}(X_i'\beta, \eps_i))^2\right]        
\end{gather*} 
\normalsize
and, by direct calculation using dominated convergence theorem, change of variables and smoothness and boundedness of $f$:
\scriptsize
\begin{align*}
\pi_{1,3}^P(X_i'\beta,\eps_i) =& E\Big[E[\mathbbm{1}\{X_i'\beta + \eps_i<X_j'\beta + \eps_j <X_k'\beta + \eps_k\}   sgn(X_k'\beta + \eps_k - 2(X_j'\beta + \eps_j) + X_i'\beta + \eps_i)  K_h(X_j'\beta + \eps_j -y) \\
& \times K_h(X_k'\beta + \eps_k -y)|X_i'\beta, X_j'\beta, X_k'\beta, \eps_i]  K_h((X_{kj}-X_{ji})'\beta)|X_i'\beta, \eps_i\Big]K_h(X_i'\beta + \eps_i - y) + \text{symmetric terms}  \\
=&  - \frac{1}{2} \mathbbm{1}\{X_i'\beta + \eps_i <y\}  E\left[f(y -X_j'\beta) f(y-X_k'\beta)K_h((X_{kj}-X_{ji})'\beta)|X_i'\beta, \eps_i\right]K_h(X_i'\beta + \eps_i - y) + \text{symmetric terms}\\
=&  sgn(X_i'\beta + \eps_i -y) E\left[f(y -X_j'\beta) f(y-X_k'\beta)K_h((X_{kj}-X_{ji})'\beta)|X_i'\beta, \eps_i\right]K_h(X_i'\beta + \eps_i - y) + o(1)\\
\equiv & sgn(X_i'\beta + \eps_i -y) a_y(X_i'\beta, \eps_i) K_h(X_i'\beta + \eps_i - y) + o(1)
\end{align*}
\normalsize
where we have also used $\int \mathbbm{1}\{s_1<s_2\} K(s_1)K(s_2) ds_1 ds_2=\frac{1}{2}$ (by symmetry of $K$) and the fact that the leading term in the sum above has an equivalent term in the sum where indices $(j,k)$ are switched.

Thus, we obtain:
\scriptsize
\begin{align*}
E[(\pi_{1,3}^P(X_i'\beta,\eps_j) - \pi_{1,3}^P(X_i'\beta,\eps_i))^2] &=  E[a_y^2(X_i'\beta, \eps_i) (sgn(X_i'\beta + \eps_j -y) K_h(X_i'\beta + \eps_j - y) - sgn(X_i'\beta + \eps_i -y)  K_h(X_i'\beta + \eps_i - y) )^2] \\
&=E[a_y^2(X_i'\beta, \eps_i) \int_{-1}^1\int_{-1}^1(sgn(s_1) K(s_1) - sgn(s_2)  K(s_2) )^2 ds_1 ds_2 f^2(y-X_i'\beta)] + o(1)\\
&=O(1)
\end{align*}
\normalsize
which implies $diam = \sup_y \sqrt{E[ \gamma_n^2(y)]} = O(\sqrt{h})$.

Let $G^{\Gamma}$ denote the envelope of $\Gamma$. We have $\|G^{\Gamma}\|_{L_2(P \times P_M)}= O(h^{-1/2})$ by the same reasoning as in \eqref{eq:condenvlp}. Now by Lemma \ref{lem:Eucl} the class $\Gamma$ is Euclidean (note that now we treat the class of functions $\{\phi_n^{\boldsymbol{\eps}}(y): y \in \mathcal{Y}\}$ unconditionally, unlike in Step 2), which implies $N(\varepsilon \|G^{\Gamma}\|_{L_2(P \times P_M)}, \Gamma, L_2(P \times P_M)) \lesssim \varepsilon^{-v}$ with $v\geq 1$ and, consequently, 
$N(\varepsilon, \Gamma, L_2(P \times P_M)) \lesssim (\sqrt{h}\varepsilon)^{-v}$. Thus, we have by Corollary 2.2.8. in \cite{van_der_vaart_wellner96}:
\begin{gather}
E  \left[\sup_y | \mathbbm{G}_P \phi_n^{\boldsymbol{\eps}}(y) -  \mathbbm{G}_P \phi_n(y)| \right] \lesssim \int_0^{O(\sqrt{h})} \sqrt{\log (\sqrt{h}\varepsilon)^{-1}} d \varepsilon = O(\sqrt{h \log h^{-1}})=o(1) \label{eq:coupling}
\end{gather}

Finally, the latter result, \eqref{eq:final1} applied under $H_0$ (i.e. with $\tilde{\boldsymbol{\eps}} = \boldsymbol{\eps}$, $f_{\tilde{\eps}} (\cdot)= f (\cdot)$ and $\tilde{P} = P$), \eqref{eq:final2} and \eqref{eq:final3} imply $|S_n^{conc} - Z|=o_p(1)$ and $|S_n^{*,conc} - Z|=o_p(1)$ which gives the final result:
\begin{gather*}
|P(S_{n}^{conc} \leq c^{conc,*}_{\alpha}) - P(Z\leq c^{conc,*}_{\alpha})|= |P(S_{n}^{conc} \leq c^{conc,*}_{\alpha}) - \alpha|+o(1)=o(1) 
\end{gather*}
(see e.g. Lemma 2.1 in \cite{chernozhukov_et_al16}).

\section{Proof of Theorem \ref{thm:bootpwr}}

Again, we focus on the local test as the argument for the global test is standard.

Note that $\inf_y U_n(y) = - \sup_y \{-U_n(y)\}$ and by applying the same argument as above we can show $\sup_y \{-U_n(y)\} = \sup_y \{-B(y)- \mathbbm{G}_P \phi_n(y) \} + o_p(1)$, which implies $\inf_y U_n(y) = \inf_y \{B(y) + \mathbbm{G}_P \phi_n(y) \} + o_p(1)$. Further, using \eqref{eq:final1} and \eqref{eq:final2} we have:
\begin{align}
\inf_y & \sqrt{nh} U_n(y) - \inf_y \sqrt{nh} U_n^*(y) \leq  \inf_y \{ \sqrt{nh} P^3 g_y^I +  \mathbbm{G}_P \phi_n(y) \} - \inf_y \{ \sqrt{nh} \tilde{P}^3 g_y^I + \mathbbm{G}_P \phi_n^{\boldsymbol{\tilde{\eps}}}(y) \} + o_p(1) \nonumber \\  
& \leq \inf_y \{ \sqrt{nh} P^3 g_y^I +  \mathbbm{G}_P \phi_n(y) \}  - \inf_y \{ \sqrt{nh} \tilde{P}^3 g_y^I + \mathbbm{G}_P \phi_n(y) \}  + o_p(1) \label{eq:pwrdiff}
\end{align}  
where the last inequality follows from \eqref{eq:coupling}.

Now by Corollary 2.2.8 in \cite{van_der_vaart_wellner96} and similar reasoning to the one leading to \eqref{eq:coupling}:
\begin{gather}
E \sup_y |\mathbbm{G}_P \phi_n(y) | \lesssim E|\mathbbm{G}_P \phi_n(0)| + K   \int_0^{O(\sqrt{h})} \sqrt{\log N(\varepsilon, \mathcal{F}^{\phi}, L_2(P)) } d \varepsilon = O(\sqrt{h \log h^{-1}}) \label{eq:Gpdiff}
\end{gather} 
where $\mathcal{F}^{\phi}=\{\phi_n(y):y \in\mathcal{Y}\}$  and by change of variables $E|\mathbbm{G}_P \phi_n(0)|  = \sqrt{h} E[|a_0(X_i'\beta,\eps_i)K_h(X_i'\beta + \eps_i)|]=O(\sqrt{h})$.

By direct calculation, using Lemma \ref{lem:int}, $T(Y) = X'\beta + \eps$: 
\footnotesize
\begin{align*}
E[g_{y}^{I}|X_j'\beta = \zeta,  X_{ji}'\beta &= \xi, X_{kj}'\beta = \xi] = \\
& = h b_K  \left[T'(y)^4 \nabla f_{\boldsymbol{\eps}}  \begin{pmatrix} T(y) -\zeta +\xi \\ T(y) -\zeta  \\ T(y) -\zeta -\xi \end{pmatrix} 
 +  T''(y)T'(y)^2 f_{\boldsymbol{\eps}} \begin{pmatrix} T(y) -\zeta +\xi \\ T(y) -\zeta   \\ T(y) -\zeta -\xi \end{pmatrix}\right]'\mathbf{1}\\
E_{\tilde{P}}[g_{y}^{I}|X_j'\beta = \zeta,  X_{ji}'\beta &= \xi, X_{kj}'\beta = \xi] = h b_K  \nabla f_{\boldsymbol{\tilde{\eps}}}  \begin{pmatrix} y -\zeta +\xi \\ y -\zeta  \\ y -\zeta -\xi \end{pmatrix} '\mathbf{1}
\end{align*}
\normalsize
where $\mathbf{1}$ is a $[1 \quad  1 \quad  1]'$ vector, $b_K$ is defined as in Lemma \ref{lem:int}, $\nabla$ denotes the gradient
and $f_{\boldsymbol{\eps}}$ denotes the joint distribution of $(\eps_i, \eps_j, \eps_k)$. Let $g(\xi)$ be the expectation of any of the expressions above conditional on $X_{ji}'\beta = X_{kj}'\beta = \xi$. We have by change of variables that $P^3 g_{y}^{I}  = E[g(\xi) f_{\xi|\xi}(\xi|\xi)] + O(h^3)$, and similarly for $\tilde{P}^3 g_{y}^{I}$, which implies:
\scriptsize
\begin{align*}
P^3 g_{y}^{I}  =&   
h b_K \int \int  \left\{ T'(y)^4 \nabla f_{\boldsymbol{\eps}}   \begin{pmatrix} T(y) -\zeta +\xi \\ T(y) -\zeta   \\ T(y) -\zeta -\xi  \end{pmatrix}' \mathbf{1}
+ T''(y)T'(y)^2 f_{\boldsymbol{\eps}} \begin{pmatrix} T(y) -\zeta +\xi \\ T(y) -\zeta  \\ T(y) -\zeta -\xi  \end{pmatrix}  \right\} 
f_{xb}(\zeta-\xi)f_{xb}(\zeta)f_{xb}(\zeta+\xi) d \zeta d\xi + O(h^3)\\
\tilde{P}^3 g_{y}^{I} = & h b_K \nabla f_{\boldsymbol{\tilde{\eps}}} \begin{pmatrix} y -\zeta +\xi \\ y -\zeta  \\ y -\zeta -\xi \end{pmatrix}  \mathbf{1}  f_{xb}(\zeta-\xi)f_{xb}(\zeta)f_{xb}(\zeta+\xi) d \zeta d\xi + O(h^3)
\end{align*}
\normalsize
where $f_{xb}$ denotes the distribution of $X_i'\beta$ and we used the fact that the joint distribution of $(X_j'\beta, X_{ji}'\beta, X_{kj}'\beta)$ at $(\zeta, \xi_1, \xi_2)$ can be written as $f_{xb}(\zeta - \xi_1)f_{xb}(\zeta)f_{xb}(\zeta+\xi_2)$. 

Note that this and \eqref{eq:Gpdiff} mean that the terms $\sqrt{nh} P^3 g_y^I$ and $\sqrt{nh} \tilde{P}^3 g_y^I$ dominate the infima in \eqref{eq:pwrdiff} and thus:
\begin{gather}
\inf_y  \sqrt{nh} U_n(y) - \inf_y \sqrt{nh} U_n^*(y) \leq \inf_y \{ \sqrt{nh} P^3 g_y^I  \}  - \inf_y \{ \sqrt{nh} \tilde{P}^3 g_y^I  \}  + o_p(1)  \label{eq:pwrfinal}
\end{gather}
Now Assumption \ref{A1}\eqref{A1a}(iii) and Lemma \ref{lem:int} give $b_K<0$. This,  $nh^3 \to \infty$ and the condition \eqref{eq:pwrcond} in Theorem \ref{thm:bootpwr} imply that the right-hand side of \eqref{eq:pwrfinal} diverges to $-\infty$, which implies the final result: 
 \begin{gather*}
P(S_n^{conc}<c^*_{\alpha})  \to 1. 
\end{gather*}

\section{Power condition for wild bootstrap}\label{app:wildboot}

Consider using wild bootstrap in Step 2 of the bootstrap procedure for the local test. Namely, we draw a random sample $\{v_i\}_{i=1}^{n}$ from a two point distribution on \{-1,1\} with $P(v_{i}=-1)=P(v_{i}=1)=1/2$ and define $\eps_{i}^{*} =v_{i} \hat{\eps}_{i}$. 

With wild bootstrap we can apply the same argument as in the proof of Theorem \ref{thm:bootpwr} but now  $\tilde{P}$  corresponds to the DGP where the dependent variable, $Y^{\#}_i$, is generated as $Y^{\#}_i = X_i'\beta+ \tilde{\tilde{\eps}}_i$ where $\tilde{\tilde{\eps}}_i$ equals $\tilde{\eps}_i$ or $-\tilde{\eps}_i$ with equal probability. Noting that $\tilde{\eps}_i = Y_i - X_i'\beta$, this means that the outcome $Y^{\#}_i = Y_i$ or $Y^{\#}_i = 2X_i'\beta - Y_i$ with equal probability, thus $f_{Y^{\#}}(y|X_i) = 0.5 f_{Y}(y|X_i) + 0.5f_{2X\beta-Y}(y|X_i)$. Thus, writing these densities in terms of the density of $\eps$ we obtain:
\scriptsize
\begin{align*}
E_{\tilde{P}}  &[g_{y}^{I} |  X_j'\beta = \zeta,   X_{ji}'\beta - \xi, X_{kj}'\beta = \xi] =\\
& =  \frac{1}{2} h a_K \Bigg[T'(y)^4 \nabla f_{\boldsymbol{\eps}}  \begin{pmatrix} T(y) -\zeta +\xi \\ T(y) -\zeta \\ T(y) -\zeta -\xi  \end{pmatrix}'\mathbf{1} - 
 \nabla f_{\boldsymbol{\eps}}  \begin{pmatrix} T(2(\zeta - \xi) - y) -\zeta +\xi \\ T(2\zeta  - y) -\zeta  \\ T(2(\zeta + \xi) - y) -\zeta -\xi  \end{pmatrix}'  \begin{pmatrix} T'(2(\zeta - \xi) - y) \\ T'(2\zeta  - y) \\  T'(2(\zeta + \xi) - y) \end{pmatrix}  \begin{pmatrix} T'(2(\zeta - \xi) - y) \\ T'(2\zeta  - y) \\  T'(2(\zeta + \xi) - y) \end{pmatrix}'\mathbf{1}\\
 &\quad +  3T''(y)T'(y)^2 f_{\boldsymbol{\eps}} \begin{pmatrix} T(y) -\zeta +\xi \\ T(y) -\zeta  \\ T(y) -\zeta -\xi  \end{pmatrix} 
  -  f_{\boldsymbol{\eps}} \begin{pmatrix} T(2(\zeta - \xi) - y) -\zeta +\xi \\ T(2\zeta  - y) -\zeta  \\ T(2(\zeta + \xi) - y) -\zeta -\xi  \end{pmatrix}  \begin{pmatrix} T''(2(\zeta - \xi) - y) T'(2\zeta  - y) T'(2(\zeta + \xi) - y)  \\ T'(2(\zeta - \xi) - y) T''(2\zeta  - y) T'(2(\zeta + \xi) - y)  \\
  T'(2(\zeta - \xi) - y) T'(2\zeta  - y) T''(2(\zeta + \xi) - y) \end{pmatrix}' \mathbf{1} \Bigg]
\end{align*}  
\normalsize
which finally gives the equivalent of condition \eqref{eq:pwrcond} for wild bootstrap:
\scriptsize
\begin{align}
\sup_y &\int  \int E_{\tilde{P}}  [g_{y}^{I} |  X_j'\beta = \zeta,   X_{ji}'\beta - \xi, X_{kj}'\beta = \xi] f_{xb}(\zeta-\xi)f_{xb}(\zeta)f_{xb}(\zeta+\xi) d \zeta d\xi \nonumber \\
&- \sup_y \int  \int  \left[ T'(y)^4  \nabla f_{\boldsymbol{\eps}}   \begin{pmatrix} T(y) -\zeta +\xi \\ T(y) -\zeta   \\ T(y) -\zeta -\xi  \end{pmatrix}'\mathbf{1} 
  + 3T''(y)T'(y)^2 f_{\boldsymbol{\eps}} \begin{pmatrix} T(y) -\zeta +\xi \\ T(y) -\zeta  \\ T(y) -\zeta -\xi  \end{pmatrix} \right] f_{xb}(\zeta-\xi)f_{xb}(\zeta)f_{xb}(\zeta+\xi) d \zeta d\xi < 0
  \label{eq:pwrcond2}
\end{align}
\normalsize

\section{Additional MC simulations}
\subsection{Semiparametric bootstrap residuals}\label{app:MCnpar}
We perform a small Monte Carlo exercise in which we change the way we generate residuals in our bootstrap procedure for the local test. Namely, instead of using OLS residuals we generate $\hat{\eps}_i = \hat{T}(Y_i) - X'\hat{\beta}$, where $\hat{T}$ is the Chen's estimator of the transformation function (\cite{chen02}) and $\hat{\beta}$ is the estimator in \cite{cavanagh_sherman98}. Here we use a grid of values for $y$: -2:0.5:2, a coarser grid than in the main text. 
\begin{table}[ht]
	\centering
\begin{tabular}{cccccccc}			
\toprule						
	&		&	\multicolumn{3}{c}{Parametric residuals}					&	\multicolumn{3}{c}{Semiparametric residuals}					\\ 
	&		&	$n=100$	&	$n=250$	&	$n=500$	&	$n=100$	&	$n=250$	&	$n=500$	 \\ \midrule
H0 true	&	D0	&	0.075	&	0.049	&	0.052	&	0.076	&	0.076	&	0.057	 \\
H0 true	&	D1	&	0.000	&	0.000	&	0.000	&	0.000	&	0.000	&	0.000	 \\
H0 false	&	D2	&	0.886	&	0.995	&	1.000	&	0.789	&	0.990	&	1.000	 \\
H0 false	&	D3	&	1.000	&	1.000	&	1.000	&	1.000	&	1.000	&	1.000	 \\
H0 false	&	D4	&	0.749	&	0.789	&	0.756	&	0.769	&	0.808	&	0.792	 \\
\bottomrule
\end{tabular}																													
		\caption{Local test, test of concavity, rejection probabilities, 5\% level}
	\label{tab:MCparvsnpar}
		\floatfoot{Note: 1000 Monte Carlo simulations, 500 bootstrap replications.} 
\end{table}

The results in Table \ref{tab:MCparvsnpar} show that this version of the test performs similarly to the main version in terms of controlling the level of the test. In terms of power, it performs better than the parametric test in design D4 but slightly worse in D2. However, the test with semiparametric residuals takes on average 2.5 times longer to compute than the main test.     

\subsection{Asymmetric errors and the local test}\label{app:MCnonsym}
Formally, our local test does not require the distribution of error terms to be symmetric. Thus, we investigate the performance of our local test with errors generated from the centred Gumbel distribution. For comparison we also include the results with Normal errors from the main text. 

\begin{table}[ht]
	\centering
\begin{tabular}{cccccccc}			
\toprule						
	&		&	\multicolumn{3}{c}{Normal errors}					&	\multicolumn{3}{c}{Gumbel errors}					\\
	&		&	$n=100$	&	$n=250$	&	$n=500$	&	$n=100$	&	$n=250$	&	$n=500$	 \\ \midrule
		&		&	\multicolumn{6}{c}{Parametric bootstrap}											\\
H0 true	&	D0	&	0.056	&	0.048	&	0.053	&	0.064	&	0.071	&	0.052	 \\
H0 true	&	D1	&	0.000	&	0.000	&	0.000	&	0.000	&	0.000	&	0.000	 \\
H0 false	&	D2	&	0.868	&	0.994	&	1.000	&	0.594	&	0.860	&	0.975	 \\
H0 false	&	D3	&	1.000	&	1.000	&	1.000	&	0.999	&	1.000	&	1.000	 \\
H0 false	&	D4	&	0.878	&	0.932	&	0.938	&	0.693	&	0.706	&	0.687	\\
	&		&	\multicolumn{6}{c}{Wild bootstrap}											\\
H0 true	&	D0	&	0.060	&	0.063	&	0.053	&	0.044	&	0.029	&	0.018	 \\
H0 true	&	D1	&	0.000	&	0.000	&	0.000	&	0.000	&	0.000	&	0.000	 \\
H0 false	&	D2	&	0.717	&	0.953	&	0.995	&	0.391	&	0.620	&	0.813	 \\
H0 false	&	D3	&	1.000	&	1.000	&	1.000	&	1.000	&	1.000	&	1.000	 \\
H0 false	&	D4	&	0.615	&	0.657	&	0.677	&	0.443	&	0.414	&	0.408	\\
\bottomrule
\end{tabular}																													
		\caption{Local test, test of concavity, rejection probabilities, 5\% level}
	\label{tab:MCnonsym}
		\floatfoot{Note: 1000 Monte Carlo simulations, 500 bootstrap replications.} 
\end{table}

Table \ref{tab:MCnonsym} shows that our main procedure provides similar size control under both symmetric and asymmetric errors, with lower power under the latter. Unsurprisingly, the wild bootstrap which imposes error symmetry in the bootstrap sample, performs poorly both in terms of size control and power under Gumbel errors.

\small
\bibliographystyle{agsm}
\bibliography{Testing_transformations_curvature251125}	

\end{document}